\documentclass[american,english,twocolumn]{IEEEtran}
\usepackage[T1]{fontenc}
\usepackage[latin9]{inputenc}
\usepackage{array}
\usepackage{verbatim}
\usepackage{booktabs}
\usepackage{amsthm}
\usepackage{amsmath}
\usepackage{amssymb}
\usepackage{graphicx}
\usepackage{esint}

\makeatletter

\providecommand{\tabularnewline}{\\}

  \theoremstyle{plain}
  \newtheorem{thm}{\protect\theoremname}
\theoremstyle{definition}
\newtheorem{assumption}{Assumption}
  \theoremstyle{remark}
  \newtheorem{rem}{\protect\remarkname}
  \theoremstyle{plain}
  \newtheorem{lem}{\protect\lemmaname}

\usepackage{cite}
\IEEEoverridecommandlockouts

\makeatother

\usepackage{babel}
\addto\captionsamerican{\renewcommand{\lemmaname}{Lemma}}
\addto\captionsamerican{\renewcommand{\remarkname}{Remark}}
\addto\captionsamerican{\renewcommand{\theoremname}{Theorem}}
\addto\captionsenglish{\renewcommand{\lemmaname}{Lemma}}
\addto\captionsenglish{\renewcommand{\remarkname}{Remark}}
\addto\captionsenglish{\renewcommand{\theoremname}{Theorem}}
\providecommand{\lemmaname}{Lemma}
\providecommand{\remarkname}{Remark}
\providecommand{\theoremname}{Theorem}

\begin{document}

\title{Efficient model-based reinforcement learning for approximate online
optimal control%
\thanks{Rushikesh Kamalapurkar, Joel A. Rosenfeld, and Warren E. Dixon are
with the Department of Mechanical and Aerospace Engineering, University
of Florida, Gainesville, FL, USA. Email: \{rkamalapurkar, joelar,
wdixon\}@ufl.edu.%
}%
\thanks{This research is supported in part by NSF award numbers 1161260 and
1217908, ONR grant number N00014-13-1-0151, and a contract with the
AFRL Mathematical \foreignlanguage{american}{Modeling} and Optimization
Institute. Any opinions, findings and conclusions or recommendations
expressed in this material are those of the authors and do not necessarily
reflect the views of the sponsoring agency.%
}}

\author{Rushikesh Kamalapurkar, Joel A. Rosenfeld, and Warren E. Dixon}
\maketitle
\begin{abstract}
In this paper the infinite horizon optimal regulation problem is solved
online for a deterministic control-affine nonlinear dynamical system
using the state following (StaF) kernel method to approximate the
value function. Unlike traditional methods that aim to approximate
a function over a large compact set, the StaF kernel method aims to
approximate a function in a small neighborhood of a state that travels
within a compact set. Simulation results demonstrate that stability
and approximate optimality of the control system can be achieved with
significantly fewer basis functions than may be required for global
approximation methods.
\end{abstract}

\section{Introduction}

Reinforcement learning (RL) has become a popular tool for determining
online solutions of optimal control problems for systems with finite
state and action spaces\cite{Sutton1998,Bertsekas2007,Mehta.Meyn2009}.
Due to technical challenges, implementation of RL in systems with
continuous state and action spaces has remained an open problem. In
recent years, adaptive dynamic programming (ADP) has been successfully
used to implement RL in deterministic autonomous control-affine systems
to solve optimal control problems via value function approximation
\cite{Doya2000,Padhi2006,Al-Tamimi2008,Lewis.Vrabie2009,Dierks2009,Mehta.Meyn2009,Vamvoudakis2010,Zhang.Cui.ea2011,Bhasin.Kamalapurkar.ea2013a,Zhang.Cui.ea2013,Zhang.Liu.ea2013}.
ADP techniques employ parametric function approximation (typically
by employing neural networks (NNs)) to approximate the value function.
Implementation of function approximation in ADP is challenging because
the controller is void of pre-designed stabilizing feedback and is
completely defined by the estimated parameters. Hence, the error between
the optimal and the estimated value function is required to decay
to a sufficiently small bound sufficiently fast to establish closed-loop
stability. The size of the error bound is determined by the selected
basis functions, and the convergence rate is determined by richness
of the data used for learning.

Sufficiently accurate approximation of the value function over a sufficiently
large neighborhood often requires a large number of basis functions,
and hence, introduces a large number of unknown parameters. %
One way to achieve accurate function approximation with fewer unknown
parameters is to use some knowledge about the system to determine
the basis functions. %
However, for general nonlinear systems, prior knowledge of the features
of the optimal value function is generally not available; hence, a
large number of generic basis functions is often the only feasible
option.

Sufficiently fast approximation of the value function over a sufficiently
large neighborhood requires sufficiently rich data to be available
for learning. In traditional ADP methods such as \cite{Vamvoudakis2009,Vamvoudakis2010,Bhasin.Kamalapurkar.ea2013a},
richness of data manifests itself as the amount of excitation in the
system. In experience replay-based techniques such as \cite{Chowdhary2010a,Chowdhary.Johnson2011,Chowdhary.Yucelen.ea2012,Modares.Lewis.ea2014},
richness of data is quantified by eigenvalues of the recorded history
stack. In model-based RL techniques such as \cite{Kamalapurkar.Walters.ea2013,Kamalapurkar.Andrews.ea2014,Kamalapurkar.Klotz.eatoappear},
richness of data corresponds to the eigenvalues of a learning matrix.
As the dimension of the system and the number of basis functions increases,
the required richness of data increases. In traditional ADP methods,
the demand for richer data causes the designer to design increasingly
aggressive excitation signals, thereby causing undesirable oscillations.
Hence, implementation of traditional ADP techniques such as \cite{Doya2000,Padhi2006,Al-Tamimi2008,Lewis.Vrabie2009,Dierks2009,Mehta.Meyn2009,Vamvoudakis2009,Vamvoudakis2010,Zhang.Cui.ea2011,Bhasin.Kamalapurkar.ea2013a,Zhang.Cui.ea2013,Zhang.Liu.ea2013}
in high dimensional systems are seldom found in the literature. In
experience replay-based ADP methods and in model-based RL, the demand
for richer data causes the required amount of data stored in the history
stack, and the number of points selected to construct the learning
matrix, respectively, to grow exponentially with the dimension of
the system. Hence, implementation of data-driven ADP techniques such
as \cite{Kamalapurkar.Walters.ea2013,Kamalapurkar.Andrews.ea2014,Modares.Lewis.ea2014,Luo.Wu.ea2014,Yang.Liu.ea2014a,Kamalapurkar.Klotz.eatoappear}
are scarcely found in the literature. %

The contribution of this paper is the development of a novel model-based
RL technique to achieve sufficient excitation without causing undesirable
oscillations and expenditure of control effort like traditional ADP
techniques and at a lower computational cost than state-of-the-art
data-driven ADP techniques. Motivated by the fact that the computational
effort required to implement ADP and the data-richness required to
achieve convergence decrease with decreasing number of basis functions,
this paper focuses on reduction of the number of basis functions used
for value function approximation. A key contribution of this paper
is the observation that online implementation of an ADP-based approximate
optimal controller does not require an estimate of the optimal value
function over the entire domain of operation of the system. Instead,
only an estimate of the slope of the value function evaluated at the
current state is required for feedback. Hence, estimation of the value
function over a small neighborhood of the current state should be
sufficient to implement an ADP-based approximate optimal controller.
Furthermore, it is reasonable to postulate that approximation of the
value function over a smaller local domain would require fewer basis
functions as opposed to approximation over the entire domain of operation. 

In this paper, reduction in the number of basis functions required
for value function approximation is achieved via selection of basis
functions that travel with the system state (referred to as state-following
(StaF) kernels) to achieve accurate approximation of the value function
over a small neighborhood of the state. The use of StaF kernel introduces
a technical challenge owing to the fact that the ideal values of the
unknown parameters corresponding to the StaF kernels are functions
of the system state. The Lyapunov-based stability analysis presented
in Section \ref{sub:StaFStability-Analysis} explicitly incorporates
this functional relationship using the result that the ideal weights
are continuously differentiable functions of the system state.

Sufficient exploration without the addition of an aggressive excitation
signal is achieved via model-based RL based on BE extrapolation \cite{Kamalapurkar.Walters.ea2013,Kamalapurkar.Andrews.ea2014}.
The computational load associated with BE extrapolation is reduced
via the selection of a single time-varying extrapolation function
instead of a large number of autonomous extrapolation functions used
in \cite{Kamalapurkar.Walters.ea2013,Kamalapurkar.Andrews.ea2014}.
Stability and convergence to optimality are obtained under a PE condition
on the extrapolated regressor. Intuitively, selection of a single
time-varying BE extrapolation function results in virtual excitation.
That is, instead of using input-output data from a persistently excited
system, the dynamic model is used to simulate persistent excitation
to facilitate parameter convergence. Simulation results are included
to demonstrate the effectiveness of the developed technique.

\section{StaF Kernel Functions\label{sec:StaF-Kernel-Functions}}

The objective in StaF-based function approximation is to maintain
good approximation of the target function in a small region of interest
in the neighborhood of a point of interest $x\in\mathbb{R}^{n}$.
In state-of-the-art online approximate control, the optimal value
function is approximated using a linear-in-the-parameters approximation
scheme, and the approximate control law drives the system along the
steepest negative gradient of the approximated value function. To
compute the controller at the current state, only the gradient of
the value function evaluated at the current state is required. Hence,
in this application, the target function is the optimal value function,
and the point of interest is the system state. 

Since the system state evolves through the state-space with time,
the region of interest for function approximation also evolves through
the state-space. The StaF technique aims to maintain a uniform approximation
of the value function over a small region around the current system
state so that the gradient of the value function at the current state,
and hence, the optimal controller at the current state, can be approximated. 

To facilitate the theoretical development, this section summarizes
key results from \cite{Rosenfeld.Kamalapurkar.easubmitted}, where
the theory of reproducing kernel Hilbert spaces (RKHSs) is used to
establish continuous differentiability of the ideal weights with respect
to the system state, and the postulate that approximation of the value
function over a small neighborhood of the current state would require
fewer basis functions is stated and proved. 

To facilitate the discussion, let $H$ be a universal RKHS over a
compact set $\chi\subset\mathbb{R}^{n}$ with a continuously differentiable
positive definite kernel $k:\chi\times\chi\to\mathbb{R}$. Let $\overline{V}^{*}:\chi\to\mathbb{R}$
be a function such that $\overline{V}^{*}\in H$ . Let $c\triangleq\left[c_{1},c_{2},\cdots c_{L}\right]^{T}\in\chi^{L}$
be a set of distinct centers, and let $\sigma:\chi\times\chi^{L}\to\mathbb{R}^{L}$
be defined as $\sigma\left(x,c\right)=\left[k\left(x,c_{1}\right),\cdots,k\left(x,c_{L}\right)\right]^{T}$.
Then, there exists a unique set of weights $W_{H}$ such that 
\[
W_{H}\left(c\right)=\arg\min_{a\in\mathbb{R}^{L}}\left\Vert a^{T}\sigma(\cdot,c)-\overline{V}^{*}\right\Vert _{H},
\]
where $\left\Vert \cdot\right\Vert _{H}$ denotes the Hilbert space
norm.

In the StaF approach, the centers are selected to follow the current
state $x,$ i.e., $c\left(x\right)\triangleq\left[c_{1}\left(x\right),c_{2}\left(x\right),\cdots c_{L}\left(x\right)\right]^{T}:\chi\to\chi^{L}.$
Since the system state evolves in time, the ideal weights are not
constant. To approximate the ideal weights using gradient-based algorithms,
it is essential that the weights change smoothly with respect to the
system state. 

Let $B_{r}\left(x\right)\subset\chi$ denote a closed ball of radius
$r$ centered at the current state $x$. Let $H_{x,r}$ denote the
restriction of the Hilbert space $H$ to $B_{r}\left(x\right)$. Then,
$H_{x,r}$ is a Hilbert space with the restricted kernel $k_{x,r}:B_{r}\left(x\right)\times B_{r}\left(x\right)\to\mathbb{R}$
defined as $k_{x,r}\left(y,z\right)=k\left(y,z\right),\:\forall\left(y,z\right)\in B_{r}\left(x\right)\times B_{r}\left(x\right)$.
The following result, first stated and proved in \cite{Rosenfeld.Kamalapurkar.easubmitted}
is stated here to motivate the use of StaF kernels. 
\begin{thm}
\cite{Rosenfeld.Kamalapurkar.easubmitted} \label{thm:StaFNumBasis}Let
$K(x,y)=e^{x^{T}y}$ be the exponential kernel function, which corresponds
to an universal RKHS, and let $\epsilon,r>0$. Then, for each $y\in\chi$,
there exists a finite number of centers, $c_{1},c_{2},...,c_{M_{y,\epsilon}}\in B_{r}(y)$
and weights $w_{1},w_{2},...,w_{M_{y,\epsilon}}$ such that 
\[
\left\Vert \overline{V}^{*}(x)-\sum_{i=1}^{M_{y,\epsilon}}w_{i}e^{x^{T}c_{i}}\right\Vert _{B_{r}(y),\infty}<\epsilon.
\]
If $p$ is an approximating polynomial that achieves the same approximation
over $B_{r}(y)$ with degree $N_{y,\epsilon}$, then an asymptotically
similar bound can be found with $M_{y,\epsilon}$ kernel functions,
where $M_{y,\epsilon}<{n+N_{y,\epsilon}+S_{y,\epsilon} \choose N_{y,\epsilon}+S_{y,\epsilon}}$
for some constant $S_{y,\epsilon}$. Moreover, $N_{y,\epsilon}$ and
$S_{y,\epsilon}$ can be bounded uniformly over $\chi$. 
\end{thm}
The Weierstrass theorem indicates that as $r$ decreases, the degree
$N_{y,\epsilon}$ of the polynomial needed to achieve the same error
$\epsilon$ over $B_{r}(y)$ decreases\cite{Lorentz.Lorentz.ea1986}.
Hence, by Theorem \ref{thm:StaFNumBasis}, approximation of a function
over a smaller domain requires a smaller number of exponential kernels.
Furthermore, provided the region of interest is small enough, the
number of kernels required to approximate continuous functions with
arbitrary accuracy can be reduced to $n+2$ where $n$ is the state
dimension. 

The following result, first stated and proved in \cite{Rosenfeld.Kamalapurkar.easubmitted}
is stated here to facilitate Lyapunov-based stability analysis of
the closed-loop system. 
\begin{thm}
\cite{Rosenfeld.Kamalapurkar.easubmitted} \label{thm:StaFContinuous}Let
the kernel function $k$ be such that the functions $k(\cdot,c)$
are $l-$times continuously differentiable for all $c\in\chi$. Let
$C$ be an ordered collection of $M$ distinct centers, $C=(c_{1},c_{2},...,c_{M})\in\chi^{M}$,
with associated ideal weights
\[
W_{H}(C)=\arg\min_{a\in R^{M}}\left\Vert \sum_{i=1}^{M}a_{i}k(\cdot,c_{i})-V(\cdot)\right\Vert _{H}.
\]
The function $W(C)$ is $l-$times continuously differentiable with
respect to each component of $C$.
\end{thm}
Thus, if the kernels are selected as functions $c_{i}:\chi\to\chi$
of the state that are $l-$times continuously differentiable, then
the ideal weight functions $W:\chi\to\mathbb{R}^{L}$ defined as $W\left(x\right)\triangleq W_{H_{x,r}}\left(c\left(x\right)\right)$
are also $l-$times continuously differentiable.

Theorem \ref{thm:StaFNumBasis} motivates the use of StaF kernels
for model-based RL, and Theorem \ref{thm:StaFContinuous} facilitates
implementation of gradient-based update laws to learn the time-varying
ideal weights in real-time. In the following, the StaF-based function
approximation approach is used to approximately solve an optimal regulation
problem online using exact model knowledge via value function approximation.
Selection of an optimal regulation problem and the assumption that
the system dynamics are known are motivated by ease of exposition.
Using a concurrent learning-based adaptive system identifier and the
state augmentation technique developed in \cite{Kamalapurkar.Andrews.ea2014},
the technique developed in this paper can be extended to a class of
trajectory tracking problems in the presence of uncertainties in the
system drift dynamics. Simulation results in Section \ref{sub:StaFSim2}
demonstrate the performance of such an extension.

\section{StaF Kernel Functions for Online Approximate Optimal Control}

\subsection{Problem Formulation}

Consider a control affine nonlinear dynamical system of the form 
\begin{equation}
\dot{x}\left(t\right)=f\left(x\left(t\right)\right)+g\left(x\left(t\right)\right)u\left(t\right),\label{eq:StaFDyn}
\end{equation}
$t\in\mathbb{R}_{\geq t_{0}}$, where $t_{0}$ denotes the initial
time, $x:\mathbb{R}_{\geq t_{0}}\to\mathbb{R}^{n}$ denotes the system
state $f:\mathbb{R}^{n}\to\mathbb{R}^{n}$ and $g:\mathbb{R}^{n}\to\mathbb{R}^{n\times m}$
denote the drift dynamics and the control effectiveness, respectively,
and $u:\mathbb{R}_{\geq0}\to\mathbb{R}^{m}$ denotes the control input.
The functions $f$ and $g$ are assumed to be locally Lipschitz continuous.
Furthermore, $f\left(0\right)=0$ and $\nabla f:\mathbb{R}^{n}\to\mathbb{R}^{n\times n}$
is continuous. In the following, the notation $\phi^{u}\left(t;t_{0},x_{0}\right)$
denotes the trajectory of the system in (\ref{eq:StaFDyn}) under
the control signal $u$ with the initial condition $x_{0}\in\mathbb{R}^{n}$
and initial time $t_{0}\in\mathbb{R}_{\geq0}$. 

The control objective is to solve the infinite-horizon optimal regulation
problem online, i.e., to design a control signal $u$ online to minimize
the cost functional 
\begin{equation}
J\left(x,u\right)\triangleq\intop_{t_{0}}^{\infty}r\left(x\left(\tau\right),u\left(\tau\right)\right)d\tau,\label{eq:StaFJ}
\end{equation}
under the dynamic constraint in (\ref{eq:StaFDyn}) while regulating
the system state to the origin. In (\ref{eq:StaFJ}), $r:\mathbb{R}^{n}\times\mathbb{R}^{m}\to\mathbb{R}_{\geq0}$
denotes the instantaneous cost defined as 
\begin{equation}
r\left(x^{o},u^{o}\right)\triangleq Q\left(x^{o}\right)+u^{o}{}^{T}Ru^{o},\label{eq:StaFr}
\end{equation}
for all $x^{o}\in\mathbb{R}^{n}$ and $u^{o}\in\mathbb{R}^{m}$, where
$Q:\mathbb{R}^{n}\to\mathbb{R}_{\geq0}$ is a positive definite function
and $R\in\mathbb{R}^{m\times m}$ is a constant positive definite
symmetric matrix. In (\ref{eq:StaFr}) and in the reminder of this
paper, the notation $\left(\cdot\right)^{o}$ is used to denote a
dummy variable.

\subsection{Exact Solution}

It is well known that since the functions $f,$ $g,$ and $Q$ are
stationary (time-invariant) and the time-horizon is infinite, the
optimal control input is a stationary state-feedback policy $u\left(t\right)=\mbox{\ensuremath{\xi}}\left(x\left(t\right)\right)$
for some function $\xi:\mathbb{R}^{n}\to\mathbb{R}^{m}$. Furthermore,
the function that maps each state to the total accumulated cost starting
from that state and following a stationary state-feedback policy,
i.e., the value function, is also a stationary function. Hence, the
optimal value function $V^{*}:\mathbb{R}^{n}\to\mathbb{R}_{\geq0}$
can be expressed as
\begin{equation}
V^{*}\left(x^{o}\right)\triangleq\inf_{u\left(\tau\right)\mid\tau\in\mathbb{R}_{\geq t}}\intop_{t}^{\infty}r\left(\phi^{u}\left(\tau;t,x^{o}\right),u\left(\tau\right)\right)d\tau,\label{eq:StaFV*def}
\end{equation}
for all $x^{o}\in\mathbb{R}^{n}$, where $U\subset\mathbb{R}^{m}$
is a compact set. Assuming an optimal controller exists, the optimal
value function can be expressed as 
\begin{equation}
V^{*}\left(x^{o}\right)\triangleq\min_{u\left(\tau\right)\mid\tau\in\mathbb{R}_{\geq t}}\intop_{t}^{\infty}r\left(\phi^{u}\left(\tau;t,x^{o}\right),u\left(\tau\right)\right)d\tau.\label{eq:StaFVstar}
\end{equation}
The optimal value function is characterized by the corresponding HJB
equation \cite{Kirk2004}
\begin{equation}
0=\min_{u^{o}\in U}\left(\nabla V\left(x^{o}\right)\left(f\left(x^{o}\right)+g\left(x^{o}\right)u^{o}\right)+r\left(x^{o},u^{o}\right)\right),\label{eq:StaFHJBmin}
\end{equation}
for all $x^{o}\in\mathbb{R}^{n},$ with the boundary condition $V\left(0\right)=0.$
Provided the HJB in (\ref{eq:StaFHJBmin}) admits a continuously differentiable
solution, it constitutes a necessary and sufficient condition for
optimality, i.e., if the optimal value function in (\ref{eq:StaFVstar})
is continuously differentiable, then it is the unique solution to
the HJB in (\ref{eq:StaFHJBmin}) \cite{Liberzon2012}. In (\ref{eq:StaFHJBmin})
and in the following development, the notation $\nabla f\left(x,y,\cdots\right)$
denotes the partial derivative of $f$ with respect to the first argument.
The optimal control policy $u^{*}:\mathbb{R}^{n}\to\mathbb{R}^{m}$
can be determined from (\ref{eq:StaFHJBmin}) as \cite{Kirk2004}
\begin{equation}
u^{*}\left(x^{o}\right)\triangleq-\frac{1}{2}R^{-1}g^{T}\left(x^{o}\right)\left(\nabla V^{*}\left(x^{o}\right)\right)^{T}.\label{eq:StaFustar}
\end{equation}

The HJB in (\ref{eq:StaFHJBmin}) can be expressed in the open-loop
form 
\begin{equation}
\nabla V^{*}\left(x^{o}\right)\left(f\left(x^{o}\right)+g\left(x^{o}\right)u^{*}\left(x^{o}\right)\right)+r\left(x^{o},u^{*}\left(x^{o}\right)\right)=0,\label{eq:StaFHJBOL}
\end{equation}
and using (\ref{eq:StaFustar}), the HJB in (\ref{eq:StaFHJBOL})
can be expressed in the closed-loop form 
\begin{multline}
-\frac{1}{4}\nabla V^{*}\left(x^{o}\right)g\left(x^{o}\right)R^{-1}g^{T}\left(x^{o}\right)\left(\nabla V^{*}\left(x^{o}\right)\right)^{T}\\
+\nabla V^{*}\left(x^{o}\right)f\left(x^{o}\right)+Q\left(x^{o}\right)=0.\label{eq:StaFHJBCL}
\end{multline}
The optimal policy can now be obtained using (\ref{eq:StaFustar})
if the HJB in (\ref{eq:StaFHJBCL}) can be solved for the optimal
value function $V^{*}$.

\subsection{Value Function Approximation}

An analytical solution of the HJB equation is generally infeasible;
hence, an approximate solution is sought. In an approximate actor-critic-based
solution, the optimal value function $V^{*}\left(x^{o}\right)$ is
replaced by a parametric estimate $\hat{V}\left(x^{o},W\right)$,
where $W\in\mathbb{R}^{L}$ denotes the vector of ideal parameters.
Replacing $V^{*}\left(x^{o}\right)$ by $\hat{V}\left(x^{o},W\right)$
in (\ref{eq:StaFustar}), an approximation to the optimal policy $u^{*}\left(x^{o}\right)$
is obtained as $\hat{u}^{o}\left(x^{o},W\right)$. The objective of
the critic is to learn the parameters $W$, and the objective of the
actor is to implement a stabilizing controller based on the parameters
learned by the critic. Motivated by the stability analysis, the actor
and the critic maintain separate estimates $\hat{W}_{a}$ and $\hat{W}_{c}$,
respectively, of the ideal parameters $W$. Substituting the estimates
$\hat{V}$ and $\hat{u}$ for $V^{*}$ and $u^{*}$ in (\ref{eq:StaFHJBOL}),
respectively, a residual error $\delta:\mathbb{R}^{n}\times\mathbb{R}^{L}\times\mathbb{R}^{L}\to\mathbb{R}$,
called the Bellman error (BE), is computed as
\begin{multline*}
\delta\left(x^{o},\hat{W}_{c},\hat{W}_{a}\right)\triangleq r\left(x^{o},\hat{u}\left(x^{o},\hat{W}_{a}\right)\right)\\
+\nabla\hat{V}\left(x^{o},\hat{W}_{c}\right)\left(f\left(x^{o}\right)+g\left(x^{o}\right)\hat{u}\left(x^{o},\hat{W}_{a}\right)\right).
\end{multline*}
 To solve the optimal control problem, the critic aims to find a set
of parameters $\hat{W}_{c}$ and the actor aims to find a set of parameters
$\hat{W}_{a}$ such that $\delta\left(x^{o},\hat{W}_{c},\hat{W}_{a}\right)=0$,
$\forall x^{o}\in\mathbb{R}^{n}$. Since an exact basis for value
function approximation is generally not available, an approximate
set of parameters that minimizes the BE is sought. 

The expression for the optimal policy in (\ref{eq:StaFustar}) indicates
that to compute the optimal action when the system is at any given
state $x^{o}\in\mathbb{R}^{n}$, one only needs to evaluate the gradient
$\nabla V^{*}$ at $x^{o}$. Hence, to compute the optimal policy
at any given state $x^{o}$, one only needs to approximate the value
function over a small neighborhood around $x^{o}$. As established
in Theorem \ref{thm:StaFNumBasis}, the number of basis functions
required to approximate the value function is smaller if the approximation
space is smaller in the sense of set containment. Hence, in this result,
instead of aiming to obtain a uniform approximation of the value function
over the entire operating domain, which might require a computationally
intractable number of basis functions, the aim is to obtain a uniform
approximation of the value function over a small neighborhood around
the current system state. 

StaF kernels are employed to achieve the aforementioned objective.
To facilitate the development, let $\chi\subset\mathbb{R}^{n}$ be
compact. Then, for all $\epsilon>0,$ there exists a function $\overline{V}^{*}=W^{T}\left(x^{o}\right)\sigma\left(x^{o},c\left(x^{o}\right)\right)\in H$
such that $\sup_{x^{o}\in\chi}\left\Vert V^{*}\left(x^{o}\right)-\overline{V}^{*}\left(x^{o}\right)\right\Vert <\epsilon$,
where $H$ is a universal RKHS, introduced in Section \ref{sec:StaF-Kernel-Functions}
and $W:\mathbb{R}^{n}\to\mathbb{R}^{L}$ denotes the ideal weight
function. In the developed StaF-based method, a small compact set
$B_{r}\left(x^{o}\right)$ around the current state $x^{o}$ is selected
for value function approximation by selecting the centers $c^{o}$
such that $c^{o}=c\left(x^{o}\right)\in B_{r}\left(x^{o}\right)$
for some function $c:\chi\to\mathbb{R}^{nL}$. The approximate value
function $\hat{V}:\chi\times\mathbb{R}^{L}\to\mathbb{R}$ and the
approximate policy $\hat{u}:\chi\times\mathbb{R}^{L}\to\mathbb{R}$
can then be expressed as
\begin{align}
\hat{V}\left(x^{o},\hat{W}_{c}\right) & \triangleq\hat{W}_{c}^{T}\sigma\left(x^{o},c\left(x^{o}\right)\right),\nonumber \\
\hat{u}\left(x^{o},\hat{W}_{a}\right) & \triangleq-\frac{1}{2}R^{-1}g^{T}\left(x^{o}\right)\nabla\sigma\left(x^{o},c\left(x^{o}\right)\right)^{T}\hat{W}_{a},\label{eq:StaFuHato}
\end{align}
where $\sigma:\chi\times\chi^{L}\to\mathbb{R}^{L}$ denotes the vector
of basis functions introduced in Section \ref{sec:StaF-Kernel-Functions}.

It should be noted that since the centers of the kernel functions
change as the system state changes, the ideal weights also change
as the system state changes. The state-dependent nature of the ideal
weights differentiates this approach from state-of-the-art ADP methods
in the sense that the stability analysis needs to account for changing
ideal weights. Based on Theorem \ref{thm:StaFContinuous}, it can
be established that the ideal weight function $W$ defined as 
\[
W\left(x\right)\triangleq\arg\min_{a\in\mathbb{R}^{L}}\left\Vert a^{T}\sigma\left(\cdot,c\left(x\right)\right)-\overline{V}^{*}\left(\cdot\right)\right\Vert _{H_{x,r}},
\]
is continuously differentiable with respect to the system state provided
the functions $\sigma$ and $c$ are continuously differentiable.

\subsection{Online Learning Based on Simulation of Experience}

To learn the ideal parameters online, the critic evaluates a form
$\delta_{t}:\mathbb{R}_{\geq t_{0}}\to\mathbb{R}$ of the BE at each
time instance $t$ as 
\begin{equation}
\delta_{t}\left(t\right)\triangleq\delta\left(x\left(t\right),\hat{W}_{c}\left(t\right),\hat{W}_{a}\left(t\right)\right),\label{eq:StaFdeltaxdot}
\end{equation}
where $\hat{W}_{a}\left(t\right)$ and $\hat{W}_{c}\left(t\right)$
denote the estimates of the actor and the critic weights, respectively,
at time $t$, and the notation $x\left(t\right)$ is used to denote
the state the system in (\ref{eq:StaFDyn}) at time $t$ when starting
from initial time $t_{0}$, initial state $x_{0}$, and under the
feedback controller 
\begin{equation}
u\left(t\right)=\hat{u}\left(x\left(t\right),\hat{W}_{a}\left(t\right)\right).\label{eq:StaFControl}
\end{equation}
Since (\ref{eq:StaFHJBOL}) constitutes a necessary and sufficient
condition for optimality, the BE serves as an indirect measure of
how close the critic parameter estimates $\hat{W}_{c}$ are to their
ideal values; hence, in RL literature, each evaluation of the BE is
interpreted as gained experience. Since the BE in (\ref{eq:StaFdeltaxdot})
is evaluated along the system trajectory, the experience gained is
along the system trajectory. 

Learning based on simulation of experience is achieved by extrapolating
the BE to unexplored areas of the state space. The critic selects
a set of functions $\left\{ x_{i}:\mathbb{R}^{n}\times\mathbb{R}_{\geq0}\to\mathbb{R}^{n}\right\} _{i=1}^{N}$
such that each $x_{i}$ maps the current state $x\left(t\right)$
to a point $x_{i}\left(x\left(t\right),t\right)\in B_{r}\left(x\left(t\right)\right)$. 

The critic then evaluates a form $\delta_{ti}:\mathbb{R}_{\geq t_{0}}\to\mathbb{R}$
of the BE for each $x_{i}$ as 
\begin{equation}
\delta_{ti}\left(t\right)=\hat{W}_{c}^{T}\left(t\right)\omega_{i}\left(t\right)+r\left(x_{i}\left(x\left(t\right),t\right),\hat{u}_{i}\left(t\right)\right),\label{eq:StaFdeltaex}
\end{equation}
where 
\begin{multline*}
\hat{u}_{i}\left(t\right)\triangleq-\frac{1}{2}R^{-1}g^{T}\left(x_{i}\left(x\left(t\right),t\right)\right)\\
\cdot\nabla\sigma\left(x_{i}\left(x\left(t\right),t\right),c\left(x\left(t\right)\right)\right)^{T}\hat{W}_{a}\left(t\right),
\end{multline*}
and 
\begin{multline*}
\omega_{i}\left(t\right)\triangleq\nabla\sigma\left(x_{i}\left(x\left(t\right),t\right),c\left(x\left(t\right)\right)\right)f\left(x_{i}\left(x\left(t\right),t\right)\right)\\
-\frac{1}{2}\nabla\sigma\left(x_{i}\left(x\left(t\right),t\right),c\left(x\left(t\right)\right)\right)g\left(x_{i}\left(x\left(t\right),t\right)\right)R^{-1}\cdot\\
g^{T}\left(x_{i}\left(x\left(t\right),t\right)\right)\nabla\sigma^{T}\left(x_{i}\left(x\left(t\right),t\right),c\left(x\left(t\right)\right)\right)\hat{W}_{a}\left(t\right).
\end{multline*}
The critic then uses the BEs from (\ref{eq:StaFdeltaxdot}) and (\ref{eq:StaFdeltaex})
to improve the estimate $\hat{W}_{c}\left(t\right)$ using the recursive
least-squares-based update law
\begin{equation}
\dot{\hat{W}}_{c}=-\eta_{c1}\Gamma\left(t\right)\frac{\omega\left(t\right)}{\rho\left(t\right)}\delta_{t}\left(t\right)-\frac{\eta_{c2}}{N}\Gamma\left(t\right)\sum_{i=1}^{N}\frac{\omega_{i}\left(t\right)}{\rho_{i}\left(t\right)}\delta_{ti}\left(t\right),\label{eq:StaFWcHD}
\end{equation}
where 
\begin{multline*}
\omega\left(t\right)\triangleq\nabla\sigma\left(x\left(t\right),c\left(x\left(t\right)\right)\right)f\left(x\left(t\right)\right)\\
-\frac{1}{2}\nabla\sigma\left(x\left(t\right),c\left(x\left(t\right)\right)\right)g\left(x\left(t\right)\right)R^{-1}g^{T}\left(x\left(t\right)\right)\\
\cdot\nabla\sigma^{T}\left(x\left(t\right),c\left(x\left(t\right)\right)\right)\hat{W}_{a}\left(t\right),
\end{multline*}
$\rho_{i}\left(t\right)\triangleq\sqrt{1+\nu\omega_{i}^{T}\left(t\right)\omega_{i}\left(t\right)}$,
$\rho\left(t\right)\triangleq\sqrt{1+\nu\omega^{T}\left(t\right)\omega\left(t\right)}$,
$\eta_{c1},\eta_{c2},\nu\in\mathbb{R}_{>0}$ are constant learning
gains, and $\Gamma\left(t\right)$ denotes the least-square learning
gain matrix updated according to
\begin{multline}
\dot{\Gamma}\left(t\right)=\beta\Gamma\left(t\right)-\eta_{c1}\Gamma\left(t\right)\frac{\omega\left(t\right)\omega^{T}\left(t\right)}{\rho^{2}\left(t\right)}\Gamma\left(t\right)\\
-\frac{\eta_{c2}}{N}\Gamma\left(t\right)\sum_{i=1}^{N}\frac{\omega_{i}\left(t\right)\omega_{i}^{T}\left(t\right)}{\rho_{i}^{2}\left(t\right)}\Gamma\left(t\right),\:\:\Gamma\left(0\right)=\Gamma_{0}.\label{eq:StaFGammD}
\end{multline}
In (\ref{eq:StaFGammD}), $\beta\in\mathbb{R}_{>0}$ is a constant
forgetting factor.

Motivated by a Lyapunov-based stability analysis, the actor improves
the estimate $\hat{W}_{a}\left(t\right)$ using the update law 
\begin{multline}
\dot{\hat{W}}_{a}\left(t\right)=-\eta_{a1}\left(\hat{W}_{a}\left(t\right)-\hat{W}_{c}\left(t\right)\right)-\eta_{a2}\hat{W}_{a}\left(t\right)+\\
\frac{\eta_{c1}G_{\sigma}^{T}\left(t\right)\hat{W}_{a}\left(t\right)\omega\left(t\right)^{T}}{4\rho\left(t\right)}\hat{W}_{c}\left(t\right)\\
+\sum_{i=1}^{N}\frac{\eta_{c2}G_{\sigma i}^{T}\left(t\right)\hat{W}_{a}\left(t\right)\omega_{i}^{T}\left(t\right)}{4N\rho_{i}\left(t\right)}\hat{W}_{c}\left(t\right),\label{eq:StaFWaHD}
\end{multline}
where $\eta_{a1},\eta_{a2}\in\mathbb{R}_{>0}$ are learning gains,
\begin{multline*}
G_{\sigma}\left(t\right)\triangleq\nabla\sigma\left(x\left(t\right),c\left(x\left(t\right)\right)\right)g\left(x\left(t\right)\right)R^{-1}g^{T}\left(x\left(t\right)\right)\\
\cdot\nabla\sigma^{T}\left(x\left(t\right),c\left(x\left(t\right)\right)\right),
\end{multline*}
and 
\begin{multline*}
G_{\sigma i}\left(t\right)\triangleq\nabla\sigma\left(x_{i}\left(x\left(t\right),t\right),c\left(x\left(t\right)\right)\right)g\left(x_{i}\left(x\left(t\right),t\right)\right)R^{-1}\\
\cdot g^{T}\left(x_{i}\left(x\left(t\right),t\right)\right)\nabla\sigma^{T}\left(x_{i}\left(x\left(t\right),t\right),c\left(x\left(t\right)\right)\right).
\end{multline*}

\section{Stability Analysis\label{sub:StaFStability-Analysis}}

For notational brevity, time-dependence of all the signals is suppressed
hereafter. Let $B_{\zeta}\subset\mathbb{R}^{n+2L}$ denote a closed
ball with radius $\zeta$ centered at the origin. Let $B_{\chi}\triangleq B_{\zeta}\cap\mathbb{R}^{n}$.
Let the notation $\overline{\left\Vert \left(\cdot\right)\right\Vert }$
be defined as $\overline{\left\Vert h\right\Vert }\triangleq\sup_{\xi\in B_{\chi}}\left\Vert h\left(\xi\right)\right\Vert $,
for some continuous function $h:\mathbb{R}^{n}\to\mathbb{R}^{k}$.
To facilitate the subsequent stability analysis, the BEs in (\ref{eq:StaFdeltaxdot})
and (\ref{eq:StaFdeltaex}) are expressed in terms of the weight estimation
errors $\tilde{W}_{c}\triangleq W-\hat{W}_{c}$ and $\tilde{W}_{a}=W-\hat{W}_{a}$
as %
\begin{align}
\delta_{t} & =-\omega^{T}\tilde{W}_{c}+\frac{1}{4}\tilde{W}_{a}G_{\sigma}\tilde{W}_{a}+\Delta\left(x\right),\nonumber \\
\delta_{ti} & =-\omega_{i}^{T}\tilde{W}_{c}+\frac{1}{4}\tilde{W}_{a}^{T}G_{\sigma i}\tilde{W}_{a}+\Delta_{i}\left(x\right).\label{eq:StaFDeltaErr}
\end{align}
where the functions $\Delta,\Delta_{i}:\mathbb{R}^{n}\to\mathbb{R}$
are uniformly bounded over $B_{\chi}$ such that the bounds $\overline{\left\Vert \Delta\right\Vert }$
and $\overline{\left\Vert \Delta_{i}\right\Vert }$ decreases with
decreasing $\overline{\left\Vert \nabla\epsilon\right\Vert }$. Let
a candidate Lyapunov function $V_{L}:\mathbb{R}^{n+2L}\times\mathbb{R}_{\geq0}\to\mathbb{R}$
be defined as 
\[
V_{L}\left(Z,t\right)\triangleq V^{*}\left(x\right)+\frac{1}{2}\tilde{W}_{c}^{T}\Gamma^{-1}\left(t\right)\tilde{W}_{c}+\frac{1}{2}\tilde{W}_{a}^{T}\tilde{W}_{a},
\]
where $V^{*}$ is the optimal value function, and 
\[
Z=\left[x^{T},\:\tilde{W}_{c}^{T},\:\tilde{W}_{a}^{T}\right]^{T}.
\]
To facilitate learning, the system states $x$ or the selected functions
$x_{i}$ are assumed to satisfy the following.
\begin{assumption}
\label{ass:StaFSimEx}There exists a positive constant $T\in\mathbb{R}_{>0}$
and nonnegative constants $\underline{c}_{1},\underline{c}_{2},$
and $\underline{c}_{3}\in\mathbb{R}_{\geq0}$ such that 
\begin{align*}
\underline{c}_{1}I_{L} & \leq\intop_{t}^{t+T}\left(\frac{\omega\left(\tau\right)\omega^{T}\left(\tau\right)}{\rho^{2}\left(\tau\right)}\right)d\tau,\:\forall t\in\mathbb{R}_{\geq0},\\
\underline{c}_{2}I_{L} & \leq\inf_{t\in\mathbb{R}_{\geq0}}\left(\frac{1}{N}\sum_{i=1}^{N}\frac{\omega_{i}\left(t\right)\omega_{i}^{T}\left(t\right)}{\rho_{i}^{2}\left(t\right)}\right),\\
\underline{c}_{3}I_{L}\leq & \frac{1}{N}\intop_{t}^{t+T}\left(\sum_{i=1}^{N}\frac{\omega_{i}\left(\tau\right)\omega_{i}^{T}\left(\tau\right)}{\rho_{i}^{2}\left(\tau\right)}\right)d\tau,\:\forall t\in\mathbb{R}_{\geq0}.
\end{align*}
Furthermore, at least one of $\underline{c}_{1},\underline{c}_{2},$
and $\underline{c}_{3}$ is strictly positive.\end{assumption}
\begin{rem}
Assumption \ref{ass:StaFSimEx} requires either the regressor $\omega$
or the regressor $\omega_{i}$ to be persistently exciting. The regressor
$\omega$ is completely determined by the system state $x$, and the
weights $\hat{W}_{a}$. Hence, excitation in $\omega$ vanishes as
the system states and the weights converge. Hence, in general, it
is unlikely that $\underline{c}_{1}>0$. However, the regressor $\omega_{i}$
depends on the functions $x_{i}$, which can be designed independent
of the system state $x$. Hence, heuristically, $\underline{c}_{3}$
can be made strictly positive if the signal $x_{i}$ contains enough
frequencies, and $\underline{c}_{2}$ can be made strictly positive
by selecting a large number of extrapolation functions. 

In previous model-based RL results such as \cite{Kamalapurkar.Walters.ea2013},
stability and convergence of the developed method relied on $\underline{c}_{2}$
being strictly positive. In the simulation example in Section \ref{sub:StaFSim1}
the extrapolation algorithm from \cite{Kamalapurkar.Walters.ea2013}
is used in the sense that large number of extrapolation functions
is selected to make $\underline{c}_{2}$ strictly positive. In this
example, the extrapolation algorithm from \cite{Kamalapurkar.Walters.ea2013}
is rendered computationally feasible by the fact that the value function
is a function of only two variables. However, the number of extrapolation
functions required to make $\underline{c}_{2}$ strictly positive
increases exponentially with increasing state dimension. Hence, implementation
of techniques such as \cite{Kamalapurkar.Walters.ea2013} is rendered
computationally infeasible in higher dimensions. In this paper, the
computational efficiency of model-based RL is improved by allowing
time-varying extrapolation functions that ensure that $\underline{c}_{3}$
is strictly positive, which can be achieved using a single extrapolation
trajectory that contains enough frequencies. The performance of the
developed extrapolation method is demonstrated in the simulation example
in Section (\ref{sub:StaFSim2}), where the value function is a function
of four variables, and a single time-varying extrapolation point is
used to improve computational efficiency instead of a large number
of fixed extrapolation functions.

The following Lemma facilitates the stability analysis by establishing
upper and lower bound on the eigenvalues of the least-squares learning
gain matrix $\Gamma$.\end{rem}
\begin{lem}
Provided Assumption \ref{ass:StaFSimEx} holds and $\lambda_{\min}\left\{ \Gamma_{0}^{-1}\right\} >0$,
the update law in (\ref{eq:StaFGammD}) ensures that the least squares
gain matrix satisfies 
\begin{equation}
\underline{\Gamma}I_{L}\leq\Gamma\left(t\right)\leq\overline{\Gamma}I_{L},\label{eq:StaFGammaBound}
\end{equation}
where $\overline{\Gamma}=\frac{1}{\min\left\{ \eta_{c1}\underline{c}_{1}+\eta_{c2}\max\left\{ \underline{c}_{2}T,\underline{c}_{3}\right\} ,\lambda_{\min}\left\{ \Gamma_{0}^{-1}\right\} \right\} e^{-\beta T}}$
and $\underline{\Gamma}=\frac{1}{\lambda_{\max}\left\{ \Gamma_{0}^{-1}\right\} +\frac{\left(\eta_{c1}+\eta_{c2}\right)}{\beta\nu}}$.
Furthermore, $\overline{\Gamma}>0$.\end{lem}
\begin{IEEEproof}
The proof closely follows the proof of \cite[Corollary 4.3.2]{Ioannou1996}.
The update law in (\ref{eq:StaFGammD}) implies that $\frac{d}{dt}\Gamma^{-1}\left(t\right)=-\beta\Gamma^{-1}\left(t\right)+\eta_{c1}\frac{\omega\left(t\right)\omega^{T}\left(t\right)}{\rho^{2}\left(t\right)}+\frac{\eta_{c2}}{N}\sum_{i=1}^{N}\frac{\omega_{i}\left(t\right)\omega_{i}^{T}\left(t\right)}{\rho_{i}^{2}\left(t\right)}.$
Hence,
\begin{align*}
\Gamma^{-1}\left(t\right) & =e^{-\beta t}\Gamma_{0}^{-1}+\eta_{c1}\intop_{0}^{t}e^{-\beta\left(t-\tau\right)}\frac{\omega\left(\tau\right)\omega^{T}\left(\tau\right)}{\rho^{2}\left(\tau\right)}d\tau\\
 & +\frac{\eta_{c2}}{N}\intop_{0}^{t}e^{-\beta\left(t-\tau\right)}\sum_{i=1}^{N}\frac{\omega_{i}\left(\tau\right)\omega_{i}^{T}\left(\tau\right)}{\rho_{i}^{2}\left(\tau\right)}d\tau
\end{align*}
To facilitate the proof, let $t<T$. Then, 
\[
\Gamma^{-1}\left(t\right)\geq e^{-\beta t}\Gamma_{0}^{-1}\geq e^{-\beta T}\Gamma_{0}^{-1}\geq\lambda_{\min}\left\{ \Gamma_{0}^{-1}\right\} e^{-\beta T}I_{L}.
\]
If $t\geq T,$ then since the integrands are positive, $\Gamma^{-1}$
can be bounded as 
\begin{align*}
\Gamma^{-1}\left(t\right) & \geq\eta_{c1}\intop_{t-T}^{t}e^{-\beta\left(t-\tau\right)}\frac{\omega\left(\tau\right)\omega^{T}\left(\tau\right)}{\rho^{2}\left(\tau\right)}d\tau\\
 & +\frac{\eta_{c2}}{N}\intop_{t-T}^{t}e^{-\beta\left(t-\tau\right)}\sum_{i=1}^{N}\frac{\omega_{i}\left(\tau\right)\omega_{i}^{T}\left(\tau\right)}{\rho_{i}^{2}\left(\tau\right)}d\tau.
\end{align*}
Hence, 
\begin{align*}
\Gamma^{-1}\left(t\right) & \geq\eta_{c1}e^{-\beta T}\intop_{t-T}^{t}\frac{\omega\left(\tau\right)\omega^{T}\left(\tau\right)}{\rho^{2}\left(\tau\right)}d\tau\\
 & +\frac{\eta_{c2}}{N}e^{-\beta T}\intop_{t-T}^{t}\sum_{i=1}^{N}\frac{\omega_{i}\left(\tau\right)\omega_{i}^{T}\left(\tau\right)}{\rho_{i}^{2}\left(\tau\right)}d\tau.
\end{align*}
Using Assumption \ref{ass:StaFSimEx}, 
\begin{align*}
\frac{1}{N}\intop_{t-T}^{t}\sum_{i=1}^{N}\frac{\omega_{i}\left(\tau\right)\omega_{i}^{T}\left(\tau\right)}{\rho_{i}^{2}\left(\tau\right)}d\tau & \geq\max\left\{ \underline{c}_{2}T,\underline{c}_{3}\right\} I_{L},\\
\intop_{t-T}^{t}\frac{\omega\left(\tau\right)\omega^{T}\left(\tau\right)}{\rho^{2}\left(\tau\right)}d\tau & \geq\underline{c}_{1}I_{L}.
\end{align*}
Hence a lower bound for $\Gamma^{-1}$ is obtained as, 
\begin{multline}
\Gamma^{-1}\left(t\right)\geq\min\Bigl\{\eta_{c1}\underline{c}_{1}+\eta_{c2}\max\left\{ \underline{c}_{2}T,\underline{c}_{3}\right\} ,\\
\lambda_{\min}\left\{ \Gamma_{0}^{-1}\right\} \Bigr\} e^{-\beta T}I_{L}.\label{eq:StaFGammaLowBound}
\end{multline}
Provided Assumption \ref{ass:StaFSimEx} holds, the lower bound in
(\ref{eq:StaFGammaLowBound}) is strictly positive. Furthermore, using
the facts that $\frac{\omega\left(t\right)\omega^{T}\left(t\right)}{\rho^{2}\left(t\right)}\leq\frac{1}{\nu}$
and $\frac{\omega_{i}\left(t\right)\omega_{i}^{T}\left(t\right)}{\rho_{i}^{2}\left(t\right)}\leq\frac{1}{\nu}$
for all $t\in\mathbb{R}_{\geq0}$, 
\begin{align*}
\Gamma^{-1}\!\left(t\right) & \leq\! e^{-\beta t}\Gamma_{0}^{-1}\!+\!\intop_{0}^{t}e^{-\beta\left(t-\tau\right)}\!\left(\!\eta_{c1}\frac{1}{\nu}+\frac{\eta_{c2}}{N}\sum_{i=1}^{N}\frac{1}{\nu}\!\right)I_{L}d\tau,\\
 & \leq\left(\lambda_{\max}\left\{ \Gamma_{0}^{-1}\right\} +\frac{\left(\eta_{c1}+\eta_{c2}\right)}{\beta\nu}\right)I_{L}.
\end{align*}
Since inverse of the lower and upper bounds on $\Gamma^{-1}$ are
the upper and lower bounds on $\Gamma$, respectively, the proof is
complete.
\end{IEEEproof}
Since the optimal value function is positive definite, (\ref{eq:StaFGammaBound})
and \cite[Lemma 4.3]{Khalil2002} can be used to show that the candidate
Lyapunov function satisfies the following bounds
\begin{equation}
\underline{v_{l}}\left(\left\Vert Z^{o}\right\Vert \right)\leq V_{L}\left(Z^{o},t\right)\leq\overline{v_{l}}\left(\left\Vert Z^{o}\right\Vert \right),\label{eq:StaFBound}
\end{equation}
for all $t\in\mathbb{R}_{\geq t_{0}}$ and for all $Z^{o}\in\mathbb{R}^{2+2L}$.
In (\ref{eq:StaFBound}), $\underline{v_{l}},\overline{v_{l}}:\mathbb{R}_{\geq0}\rightarrow\mathbb{R}_{\geq0}$
are class $\mathcal{K}$ functions. To facilitate the analysis, let
$\underline{c}\in\mathbb{R}_{>0}$ be a constant defined as 
\begin{equation}
\underline{c}\triangleq\frac{\beta}{2\overline{\Gamma}\eta_{c2}}+\frac{\underline{c}_{2}}{2},\label{eq:StaFmineig}
\end{equation}
and let $\iota\in\mathbb{R}_{>0}$ be a constant defined as
\begin{align*}
\iota & \triangleq\frac{3\left(\frac{\left(\eta_{c1}+\eta_{c2}\right)\overline{\left\Vert \Delta\right\Vert }}{\sqrt{v}}+\frac{\overline{\left\Vert \nabla Wf\right\Vert }}{\underline{\Gamma}}+\frac{\overline{\left\Vert \Gamma^{-1}G_{W\sigma}W\right\Vert }}{2}\right)^{2}}{4\eta_{c2}\underline{c}}\\
 & +\frac{1}{\left(\eta_{a1}+\eta_{a2}\right)}\Biggl(\frac{\overline{\left\Vert G_{W\sigma}W\right\Vert }+\overline{\left\Vert G_{V\sigma}\right\Vert }}{2}+\eta_{a2}\overline{\left\Vert W\right\Vert }\\
 & +\overline{\left\Vert \nabla Wf\right\Vert }+\frac{\left(\eta_{c1}+\eta_{c2}\right)\overline{\left\Vert G_{\sigma}\right\Vert }\overline{\left\Vert W\right\Vert }^{2}}{4\sqrt{v}}\Biggr)^{2}\\
 & +\frac{1}{2}\overline{\left\Vert G_{V\epsilon}\right\Vert }.
\end{align*}
Let $v_{l}:\mathbb{R}_{\geq0}\to\mathbb{R}_{\geq0}$ be a class $\mathcal{K}$
function such that 
\[
v_{l}\left(\left\Vert Z\right\Vert \right)\leq\frac{Q\left(x\right)}{2}+\frac{\eta_{c2}\underline{c}}{6}\left\Vert \tilde{W}_{c}\right\Vert ^{2}+\frac{\left(\eta_{a1}+\eta_{a2}\right)}{8}\left\Vert \tilde{W}_{a}\right\Vert ^{2}.
\]
The sufficient conditions for the subsequent Lyapunov-based stability
analysis are given by
\begin{align}
\frac{\eta_{c2}\underline{c}}{3} & \geq\frac{\left(\frac{\overline{\left\Vert G_{W\sigma}\right\Vert }}{2\underline{\Gamma}}+\frac{\left(\eta_{c1}+\eta_{c2}\right)\overline{\left\Vert W^{T}G_{\sigma}\right\Vert }}{4\sqrt{v}}+\eta_{a1}\right)^{2}}{\left(\eta_{a1}+\eta_{a2}\right)},\nonumber \\
\frac{\left(\eta_{a1}+\eta_{a2}\right)}{4} & \geq\left(\frac{\overline{\left\Vert G_{W\sigma}\right\Vert }}{2}+\frac{\left(\eta_{c1}+\eta_{c2}\right)\overline{\left\Vert W\right\Vert }\overline{\left\Vert G_{\sigma}\right\Vert }}{4\sqrt{v}}\right),\nonumber \\
v_{l}^{-1}\left(\iota\right) & <\overline{v_{l}}^{-1}\left(\underline{v_{l}}\left(\zeta\right)\right).\label{eq:StaFSuffCond}
\end{align}
Note that the sufficient conditions can be satisfied provided the
points for BE extrapolation are selected such that the minimum eigenvalue
$\underline{c}$, introduced in (\ref{eq:StaFmineig}) is large enough
and that the StaF kernels for value function approximation are selected
such that $\overline{\left\Vert \epsilon\right\Vert }$ and $\overline{\left\Vert \nabla\epsilon\right\Vert }$
are small enough. To improve computational efficiency, the size of
the domain around the current state where the StaF kernels provide
good approximation of the value function is desired to be small. Smaller
approximation domain results in almost identical extrapolated points,
which in turn, results in smaller $\underline{c}$. Hence, the approximation
domain cannot be selected to be arbitrarily small and needs to be
large enough to meet the sufficient conditions in (\ref{eq:StaFSuffCond}).
\begin{thm}
Provided Assumption \ref{ass:StaFSimEx} holds and the sufficient
gain conditions in (\ref{eq:StaFSuffCond}) are satisfied, the controller
in (\ref{eq:StaFControl}) and the update laws in (\ref{eq:StaFWcHD})
- (\ref{eq:StaFWaHD}) ensure that the state $x$ and the weight estimation
errors $\tilde{W}_{c}$ and $\tilde{W}_{a}$ are ultimately bounded.\end{thm}
\begin{IEEEproof}
The time-derivative of the Lyapunov function is given by
\begin{multline*}
\dot{V}_{L}=\dot{V}^{*}+\tilde{W}_{c}^{T}\Gamma^{-1}\left(\dot{W}-\dot{\hat{W}}_{c}\right)+\frac{1}{2}\tilde{W}_{c}^{T}\dot{\Gamma}^{-1}\tilde{W}_{c}\\
+\tilde{W}_{a}^{T}\left(\dot{W}-\dot{\hat{W}}_{a}\right).
\end{multline*}
Using Theorem \ref{thm:StaFContinuous}, the time derivative of the
ideal weights can be expressed as 
\begin{equation}
\dot{W}=\nabla W\left(x\right)\left(f\left(x\right)+g\left(x\right)u\right).\label{eq:StaFWDot}
\end{equation}
Using (\ref{eq:StaFWcHD}) - (\ref{eq:StaFDeltaErr}) and (\ref{eq:StaFWDot}),
the time derivative of the Lyapunov function is expressed as 
\begin{align*}
\dot{V}_{L} & =\nabla V^{*}\left(x\right)\left(f\left(x\right)+g\left(x\right)u\right)\\
 & +\tilde{W}_{c}^{T}\Gamma^{-1}\nabla W\left(x\right)\left(f\left(x\right)+g\left(x\right)u\right)\\
 & -\!\tilde{W}_{c}^{T}\Gamma^{-1}\!\left(\!-\eta_{c1}\Gamma\frac{\omega}{\rho}\!\left(\!-\omega^{T}\tilde{W}_{c}\!+\!\frac{1}{4}\tilde{W}_{a}G_{\sigma}\tilde{W}_{a}\!+\!\Delta\!\left(x\right)\!\right)\!\right)\\
 & -\tilde{W}_{c}^{T}\Gamma^{-1}\left(-\frac{\eta_{c2}}{N}\Gamma\sum_{i=1}^{N}\frac{\omega_{i}}{\rho_{i}}\frac{1}{4}\tilde{W}_{a}^{T}G_{\sigma i}\tilde{W}_{a}\right)\\
 & -\tilde{W}_{c}^{T}\Gamma^{-1}\left(-\frac{\eta_{c2}}{N}\Gamma\sum_{i=1}^{N}\frac{\omega_{i}}{\rho_{i}}\left(-\omega_{i}^{T}\tilde{W}_{c}+\Delta_{i}\left(x\right)\right)\right)
\end{align*}
\begin{align*}
 & -\frac{1}{2}\tilde{W}_{c}^{T}\Gamma^{-1}\left(\beta\Gamma-\eta_{c1}\Gamma\frac{\omega\omega^{T}}{\rho}\Gamma\right)\Gamma^{-1}\tilde{W}_{c}\\
 & -\frac{1}{2}\tilde{W}_{c}^{T}\Gamma^{-1}\left(-\frac{\eta_{c2}}{N}\Gamma\sum_{i=1}^{N}\frac{\omega_{i}\omega_{i}^{T}}{\rho_{i}}\Gamma\right)\Gamma^{-1}\tilde{W}_{c}\\
 & +\tilde{W}_{a}^{T}\left(\nabla W\left(x\right)\left(f\left(x\right)+g\left(x\right)u\right)-\dot{\hat{W}}_{a}\right).
\end{align*}
Provided the sufficient conditions in (\ref{eq:StaFSuffCond}) hold,
the time derivative of the candidate Lyapunov function can be bounded
as %
\begin{equation}
\dot{V}_{L}\leq-v_{l}\left(\left\Vert Z\right\Vert \right),\quad\forall\zeta>\left\Vert Z\right\Vert >v_{l}^{-1}\left(\iota\right).\label{eq:StaFVdot}
\end{equation}
Using (\ref{eq:StaFBound}), (\ref{eq:StaFSuffCond}), and (\ref{eq:StaFVdot}),
\cite[Theorem 4.18]{Khalil2002} can be invoked to conclude that $Z$
is ultimately bounded, in the sense that $\lim\sup_{t\to\infty}\left\Vert Z\left(t\right)\right\Vert \leq\underline{v_{l}}^{-1}\left(\overline{v_{l}}\left(\iota\right)\right).$
\end{IEEEproof}

\section{Simulation}

\subsection{\label{sub:StaFSim1}Optimal regulation problem with exact model
knowledge}

\subsubsection{Simulation parameters}

To demonstrate the effectiveness of the StaF kernels, simulations
are performed on a two-dimensional nonlinear dynamical system. The
system dynamics are given by (\ref{eq:StaFDyn}), where $x^{o}=[x_{1}^{o},\: x_{2}^{o}]^{T}$,
\begin{gather}
f\left(x^{o}\right)=\left[\begin{array}{c}
-x_{1}^{o}+x_{2}^{o}\\
-\frac{1}{2}x_{1}^{o}-\frac{1}{2}x_{2}^{o}\left(\cos\left(2x_{1}^{o}\right)+2\right)^{2}
\end{array}\right],\nonumber \\
g\left(x^{o}\right)=\left[\begin{array}{c}
0\\
\cos\left(2x_{1}^{o}\right)+2
\end{array}\right].\label{eq:StaFSymdyn}
\end{gather}
The control objective is to minimize the cost 
\begin{equation}
\intop_{0}^{\infty}\left(x^{T}\left(\tau\right)x\left(\tau\right)+u^{2}\left(\tau\right)\right)d\tau.\label{eq:StafSymCost}
\end{equation}
The system in (\ref{eq:StaFSymdyn}) and the cost in (\ref{eq:StafSymCost})
are selected because the corresponding optimal control problem has
a known analytical solution. The optimal value function is $V^{*}\left(x^{o}\right)=\frac{1}{2}x_{1}^{o2}+x_{2}^{o2}$,
and the optimal control policy is $u^{*}(x^{o})=-(cos(2x_{1}^{o})+2)x_{2}^{o}$
(cf. \cite{Vamvoudakis2010}).

To apply the developed technique to this problem, the value function
is approximated using three exponential StaF kernels, i.e, $\sigma\left(x^{o},c^{o}\right)=[\sigma_{1}\left(x^{o},c_{1}^{o}\right),\:\sigma_{2}\left(x^{o},c_{2}^{o}\right),\:\sigma_{3}\left(x^{o},c_{3}^{o}\right)]^{T}$.
The kernels are selected to be $\sigma_{i}\left(x^{o},c_{i}^{o}\right)=e^{x^{oT}c_{i}^{o}}-1,$
$i=1,\cdots,3$. The centers $c_{i}^{o}$ are selected to be on the
vertices of a shrinking equilateral triangle around the current state,
i.e., $c_{i}^{o}=x^{o}+d_{i}\left(x^{o}\right),$ $i=1,\cdots,3$,
where $d_{1}\left(x^{o}\right)=0.7\nu^{o}\left(x^{o}\right)\cdot[0,\:1]^{T}$,
$d_{2}\left(x^{o}\right)=0.7\nu^{o}\left(x^{o}\right)\cdot[0.87,\:-0.5]^{T}$,
and $d_{3}\left(x^{o}\right)=0.7\nu^{o}\left(x^{o}\right)\cdot[-0.87,\:-0.5]^{T}$,
and $\nu^{o}\left(x^{o}\right)\triangleq\left(\frac{x^{oT}x^{o}+0.01}{1+\nu_{2}x^{oT}x^{o}}\right)$
denotes the shrinking function. The point for BE extrapolation is
selected at random from a uniform distribution over a $2.1\nu^{o}\left(x\left(t\right)\right)\times2.1\nu^{o}\left(x\left(t\right)\right)$
square centered at the current state $x\left(t\right)$ so that the
function $x_{i}$ is of the form $x_{i}\left(x^{o},t\right)=x^{o}+a_{i}\left(t\right)$
for some $a_{i}\left(t\right)\in\mathbb{R}^{2}$.

The system is initialized at the initial conditions
\begin{gather*}
x\left(0\right)=[-1,\:1]^{T},\:\:\hat{W}_{c}\left(0\right)=0.4\times\mathbf{1}_{3\times1},\\
\Gamma\left(0\right)=500I_{3},\:\:\hat{W}_{a}\left(0\right)=0.7\hat{W}_{c}\left(0\right),
\end{gather*}
where $I_{3}$ denotes a $3\times3$ identity matrix and $\mathbf{1}_{3\times1}$
denotes a $3\times1$ matrix of ones. and the learning gains are selected
as 
\begin{gather*}
\eta_{c1}=0.001,\:\eta_{c2}=0.25,\:\eta_{a1}=1.2,\:\eta_{a2}=0.01,\\
\beta=0.003,\: v=0.05,\:\nu_{2}=1.
\end{gather*}

\subsubsection{Results}

\begin{figure}
\begin{centering}
\includegraphics[width=1\columnwidth]{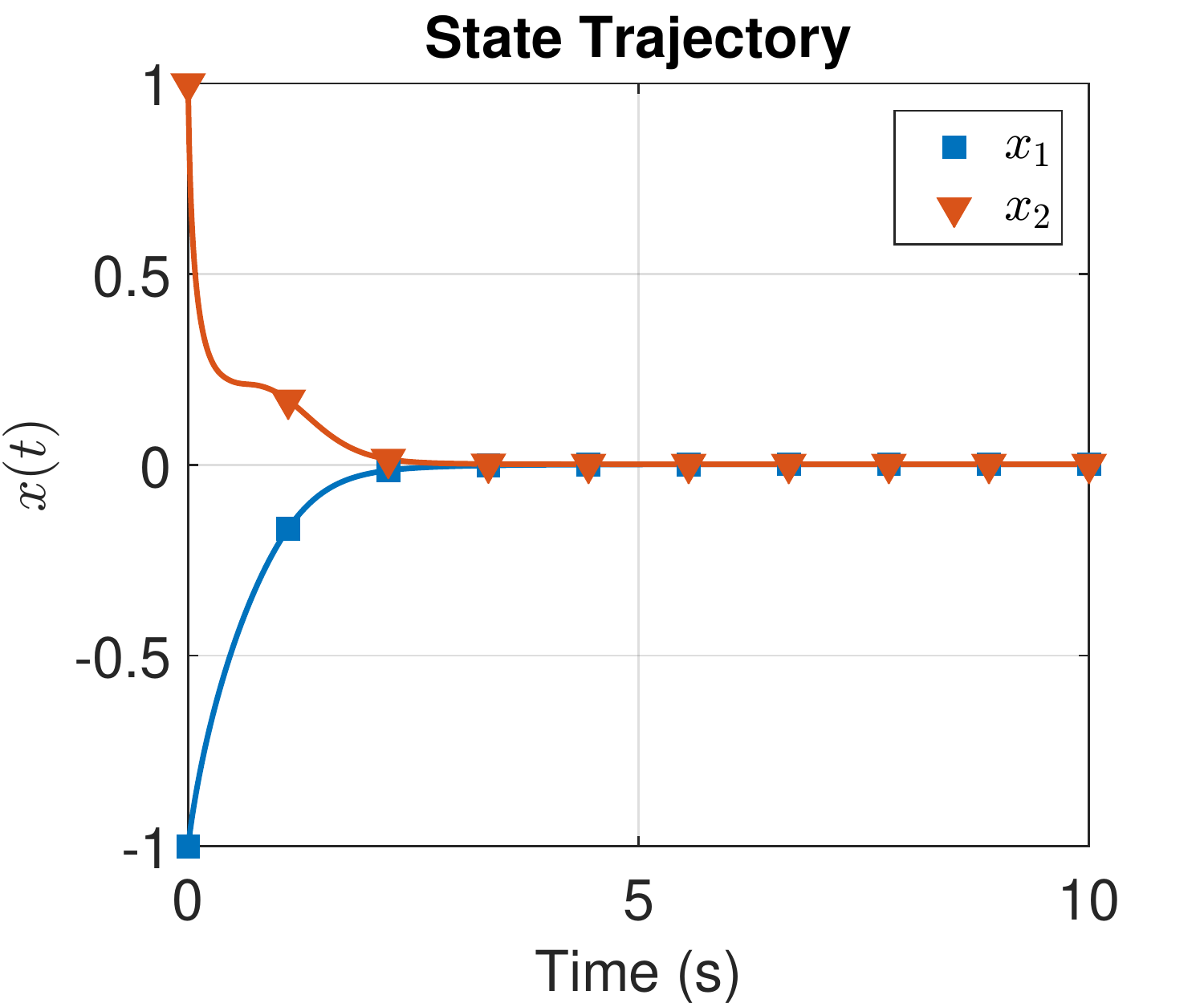}
\par\end{centering}

\protect\caption{\label{fig:StaFState}State trajectories generated using StaF kernel-based
ADP.}
\end{figure}
\begin{figure}
\begin{centering}
\includegraphics[width=1\columnwidth]{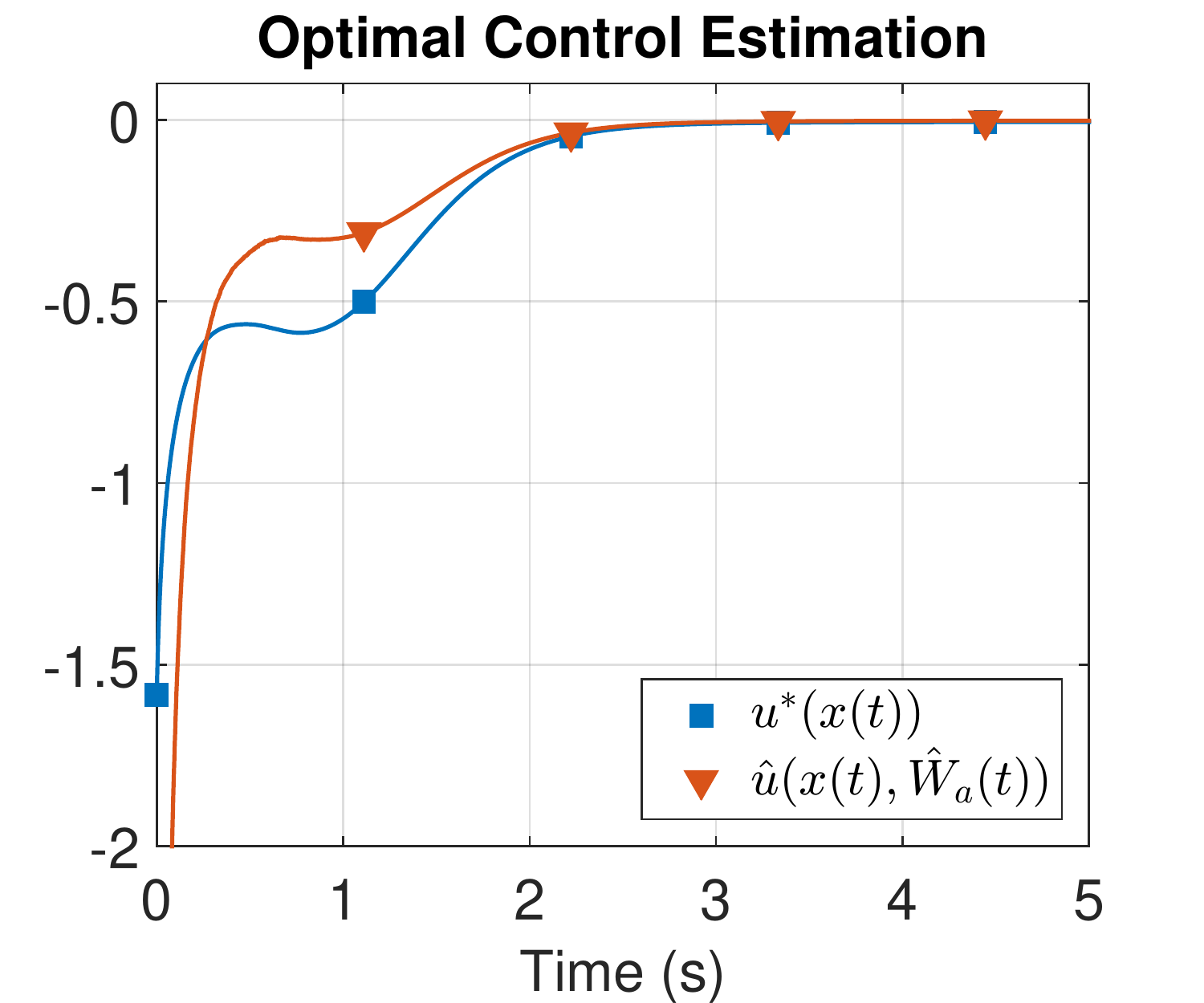}
\par\end{centering}

\protect\caption{\label{fig:StaFControl}Control trajectory generated using StaF kernel-based
ADP compared with the optimal control trajectory.}
\end{figure}
\begin{figure}
\begin{centering}
\includegraphics[width=1\columnwidth]{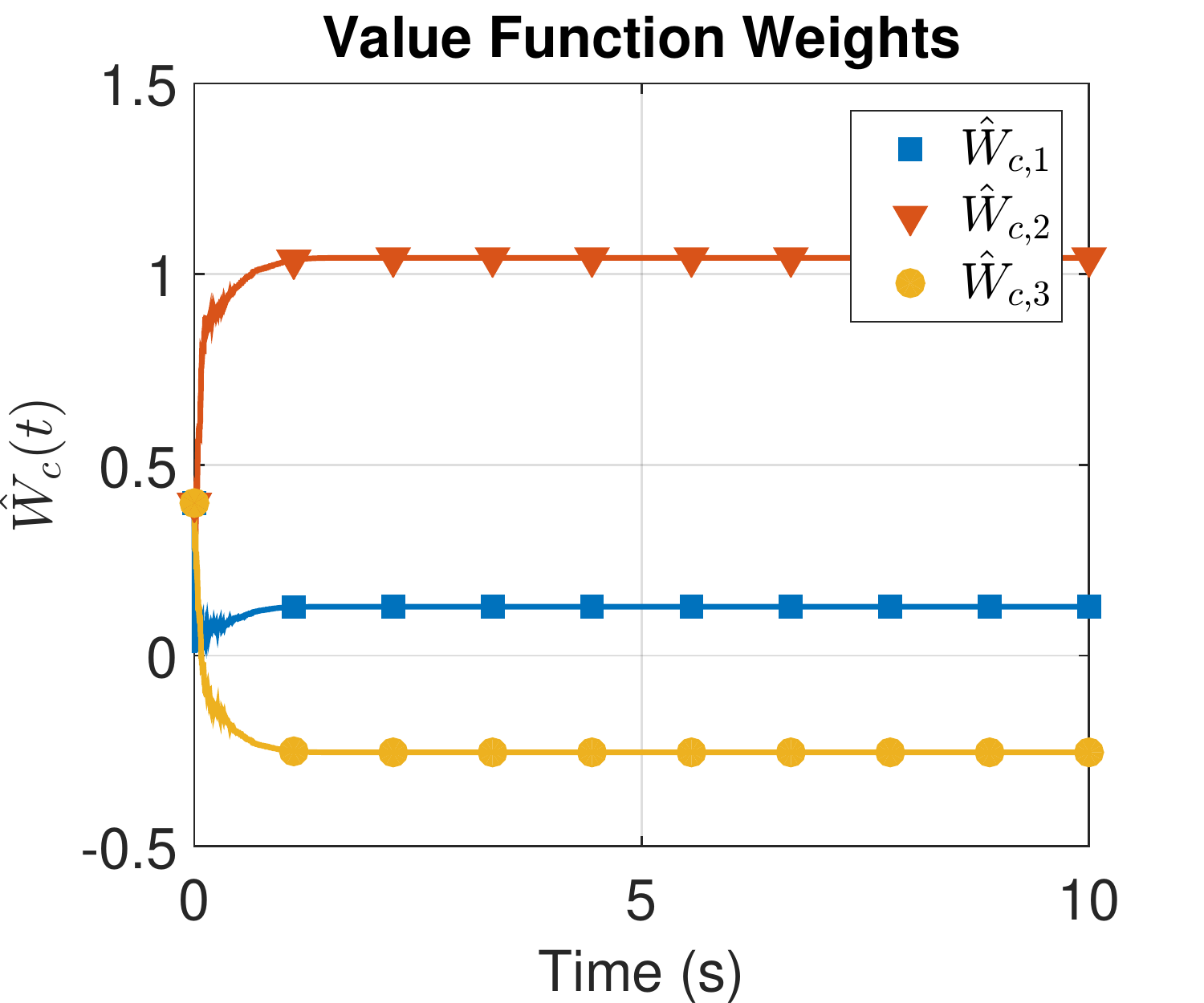}
\par\end{centering}

\protect\caption{\label{fig:StaFWc}Trajectories of the estimates of the unknown parameters
in the value function generated using StaF kernel-based ADP. The ideal
weights are unknown and time-varying; hence, the obtained weights
can not be compared with their ideal weights.}
\end{figure}
\begin{figure}
\begin{centering}
\includegraphics[width=1\columnwidth]{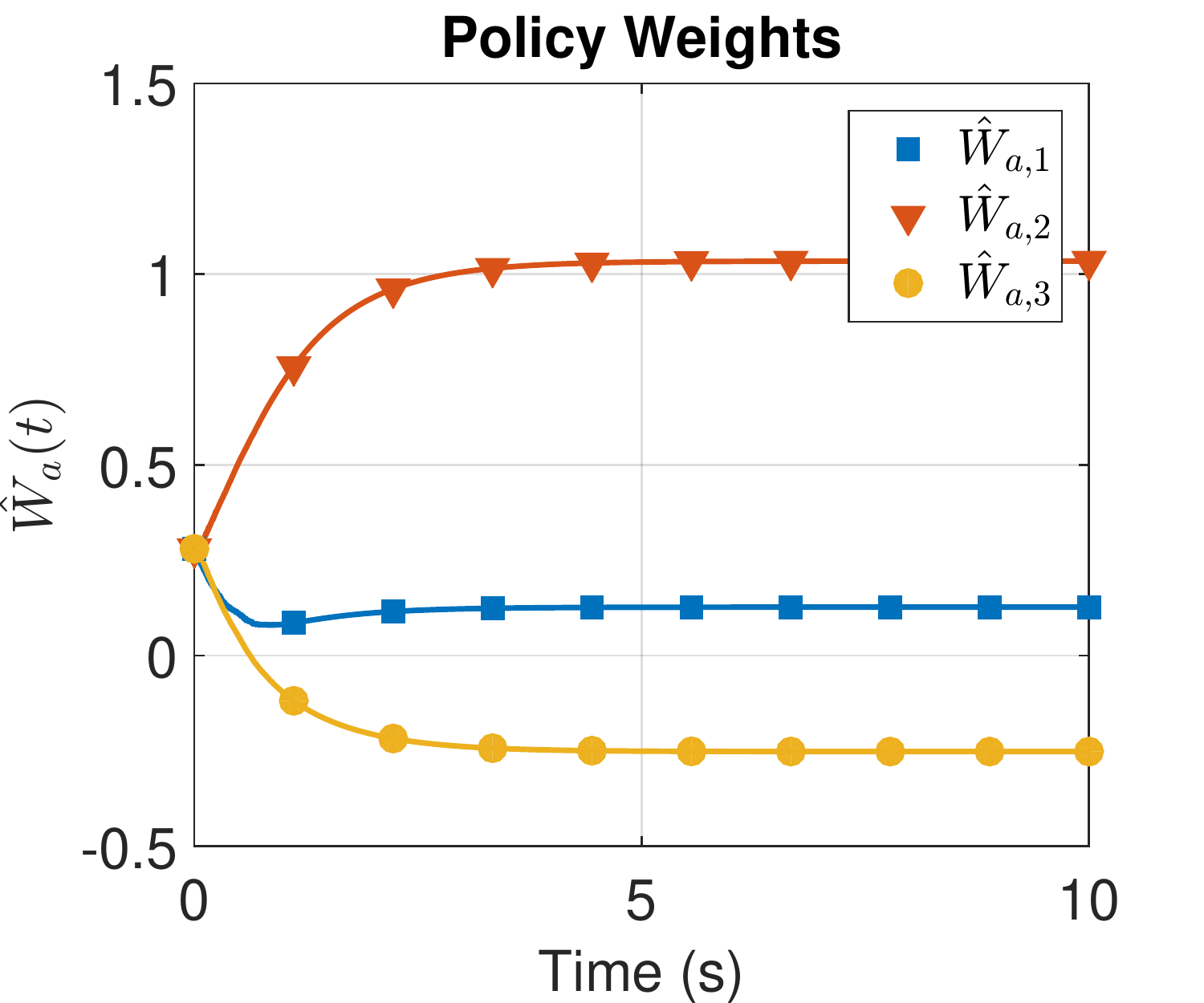}
\par\end{centering}

\protect\caption{\label{fig:StaFWa}Trajectories of the estimates of the unknown parameters
in the policy generated using StaF kernel-based ADP. The ideal weights
are unknown and time-varying; hence, the obtained weights can not
be compared with their ideal weights.}
\end{figure}
\begin{figure}
\begin{centering}
\includegraphics[width=1\columnwidth]{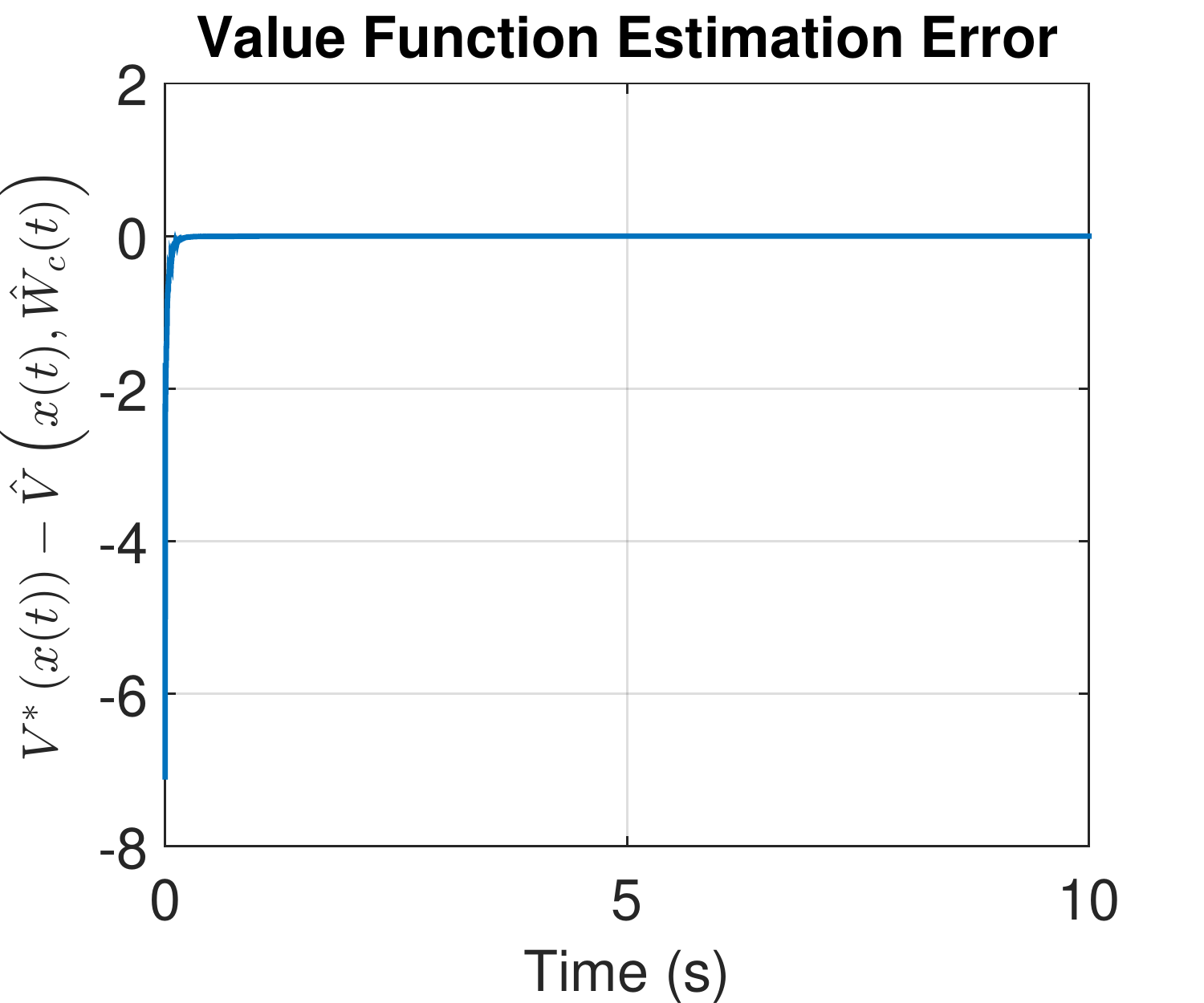}
\par\end{centering}

\protect\caption{\label{fig:StaFErrors}The error between the optimal and the estimated
value function.}
\end{figure}
Figure \ref{fig:StaFState} shows that the developed StaF-based controller
drives the system states to the origin while maintaining system stability.
Figure \ref{fig:StaFControl} shows the implemented control signal
compared with the optimal control signal. It is clear that the implemented
control converges to the optimal controller. Figure \ref{fig:StaFWc}
shows that the weight estimates for the StaF-based value function
and policy approximation remain bounded and converge as the state
converges to the origin. Since the ideal values of the weights are
unknown, the weights can not directly be compared with their ideal
values. However, since the optimal solution is known, the value function
estimate corresponding to the weights in Figure \ref{fig:StaFWc}
can be compared to the optimal value function at each time $t$. Figure
\ref{fig:StaFErrors} shows that the error between the optimal and
the estimated value functions rapidly decays to zero.

\subsection{\label{sub:StaFSim2}Optimal tracking problem with parametric uncertainties
in the drift dynamics}

\subsubsection{Simulation parameters}

Similar to \cite{Kamalapurkar.Andrews.ea2014}, the developed StaF-based
RL technique is extended to solve optimal tracking problems with parametric
uncertainties in the drift dynamics. The drift dynamics in the two-dimensional
nonlinear dynamical system in (\ref{eq:StaFSymdyn}) are assumed to
be linearly parameterized as 
\[
f\left(x^{o}\right)=\underset{\theta^{T}}{\underbrace{\left[\begin{array}{ccc}
\theta_{1} & \theta_{2} & \theta_{3}\\
\theta_{4} & \theta_{5} & \theta_{6}
\end{array}\right]}}\underset{\sigma_{\theta}\left(x^{o}\right)}{\underbrace{\left[\begin{array}{c}
x_{1}^{o}\\
x_{2}^{o}\\
x_{2}^{o}\left(cos\left(2x_{1}^{o}\right)+2\right)
\end{array}\right]}},
\]
where $\theta\in\mathbb{R}^{3\times2}$ is the matrix of unknown parameters
and $\sigma_{\theta}$ is the known vector of basis functions. The
ideal values of the unknown parameters are $\theta_{1}=-1$, $\theta_{2}=1$,
$\theta_{3}=0$, $\theta_{4}=-0.5$, $\theta_{5}=0$, and $\theta_{6}=-0.5$.
Let $\hat{\theta}$ denote an estimate of the unknown matrix $\theta.$
The control objective is to drive the estimate $\hat{\theta}$ to
the ideal matrix $\theta$, and to drive the state $x$ to follow
a desired trajectory $x_{d}$. The desired trajectory is selected
to be solution of the initial value problem 
\begin{equation}
\dot{x}_{d}\left(t\right)=\begin{bmatrix}-1 & 1\\
-2 & 1
\end{bmatrix}x_{d}\left(t\right),\quad x_{d}\left(0\right)=\begin{bmatrix}0\\
1
\end{bmatrix},\label{eq:StaFSymDesDyn}
\end{equation}
and the cost functional is selected to be $\intop_{0}^{\infty}\left(e^{T}\left(t\right)\mbox{diag}\left(10,\:10\right)e\left(t\right)+\left(\mu\left(t\right)\right)^{2}\right)dt$,
where $e\left(t\right)=x\left(t\right)-x_{d}\left(t\right),$ $\mu\left(t\right)=u\left(t\right)-g^{+}\left(x_{d}\left(t\right)\right)\left(\begin{bmatrix}-1 & 1\\
-2 & 1
\end{bmatrix}x_{d}\left(t\right)-f\left(x_{d}\left(t\right)\right)\right)$, and $g^{+}\left(x^{o}\right)$ denotes the pseudoinverse of $g\left(x^{o}\right)$.

The value function is a function of the concatenated state $\zeta\triangleq\begin{bmatrix}e^{T} & x_{d}^{T}\end{bmatrix}^{T}\in\mathbb{R}^{4}$.
The value function is approximated using five exponential StaF kernels
given by $\sigma_{i}\left(\zeta^{o},c_{i}^{o}\right)$, where the
five centers are selected according to $c_{i}^{o}\left(\zeta^{o}\right)=\zeta^{o}+d_{i}\left(\zeta^{o}\right)$
to form a regular five dimensional simplex around the current state
with $\nu^{o}\left(\zeta^{o}\right)\equiv1$. Learning gains for system
identification and value function approximation are selected as
\begin{gather*}
\mbox{\ensuremath{\eta}}_{c1}=0.001,\:\eta_{c2}=2,\:\eta_{a1}=2,\:\eta_{a2}=0.001,\\
\beta=0.01,\:\nu=0.1,\:\nu_{2}=1,\: k=500,\\
\mbox{\ensuremath{\Gamma}}_{\theta}=I_{3},\:\Gamma\left(0\right)=50I_{5},\: k_{\theta}=20,
\end{gather*}
To implement BE extrapolation, a single state trajectory $\zeta_{i}$
is selected as $\zeta_{i}\left(\zeta^{o},t\right)=\zeta^{o}+a_{i}\left(t\right)$,
where $a_{i}\left(t\right)$ is sampled at each $t$ from a uniform
distribution over the a $2.1\times2.1\times2.1\times2.1$ hypercube
centered at the origin. The history stack required for CL contains
ten points, and is recorded online using a singular value maximizing
algorithm (cf. \cite{Chowdhary.Yucelen.ea2012}), and the required
state derivatives are computed using a fifth order Savitzky-Golay
smoothing filter (cf. \cite{Savitzky.Golay1964}).

The initial values for the state and the state estimate are selected
to be $x\left(0\right)=[0,0]^{T}$ and $\hat{x}\left(0\right)=[0,0]^{T}$,
respectively. The initial values for the NN weights for the value
function, the policy, and the drift dynamics are selected to be $0.025\times\mathbf{1}_{5}$,
$0.025\times\mathbf{1}_{5}$, and $\mathbf{0}_{3\times2}$, respectively,
where $\mathbf{0}_{3\times2}$ denotes a $3\times2$ matrix of zeros.
Since the system in (\ref{eq:StaFSymdyn}) has no stable equilibria,
the initial policy $\hat{\mu}\left(\zeta,\mathbf{0}_{3\times2}\right)$
is not stabilizing. The stabilization demonstrated in Figure \ref{fig:StaFTracking-error-trajectories}
is achieved via fast simultaneous learning of the system dynamics
and the value function.

\subsubsection{Results}

\begin{figure}[h]
\noindent \begin{centering}
\includegraphics[width=1\columnwidth]{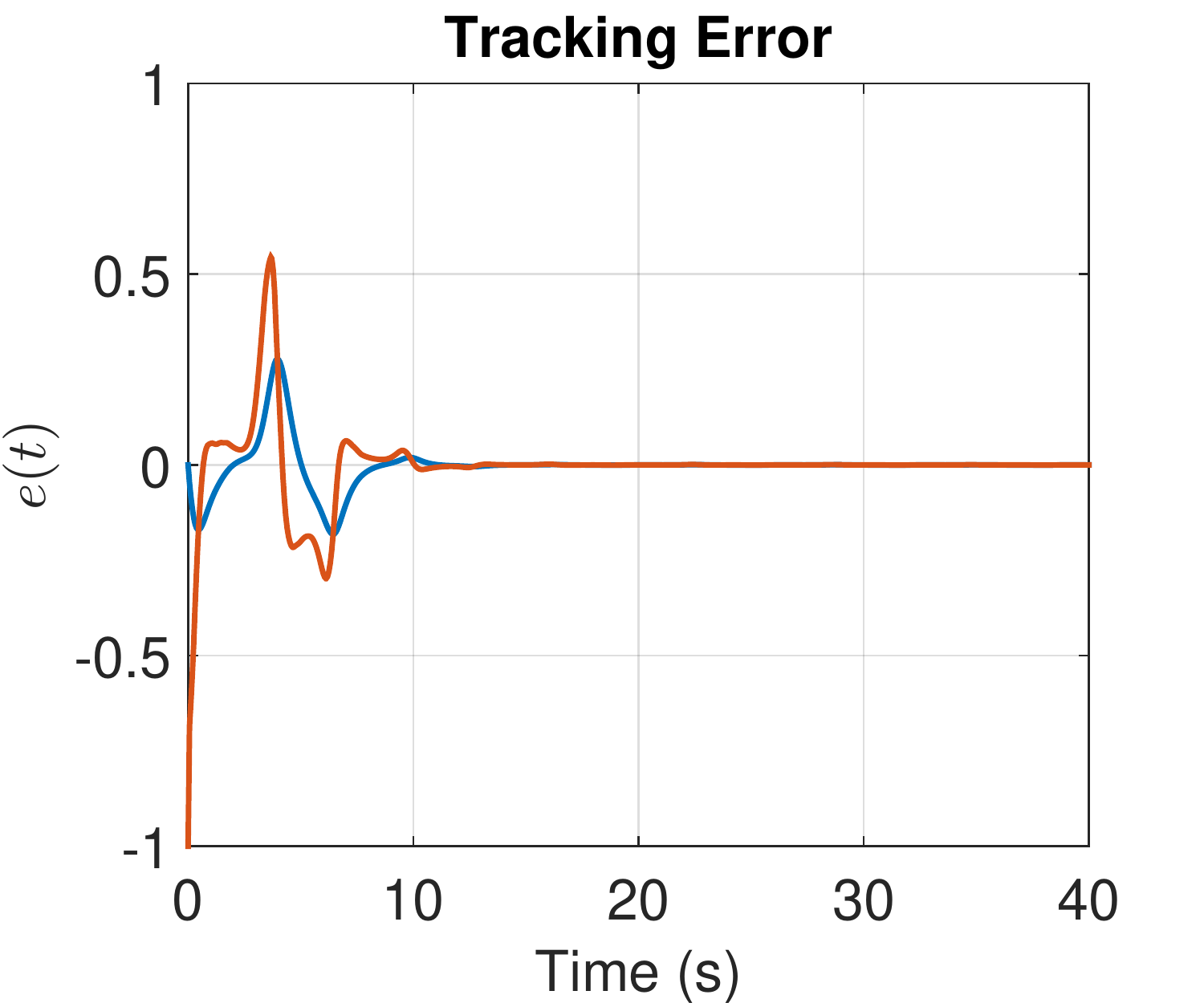}
\par\end{centering}

\noindent \centering{}\protect\caption{\label{fig:StaFTracking-error-trajectories}Tracking error trajectories
generated using the proposed method for the nonlinear system.}
\end{figure}

\begin{figure}[h]
\noindent \begin{centering}
\includegraphics[width=1\columnwidth]{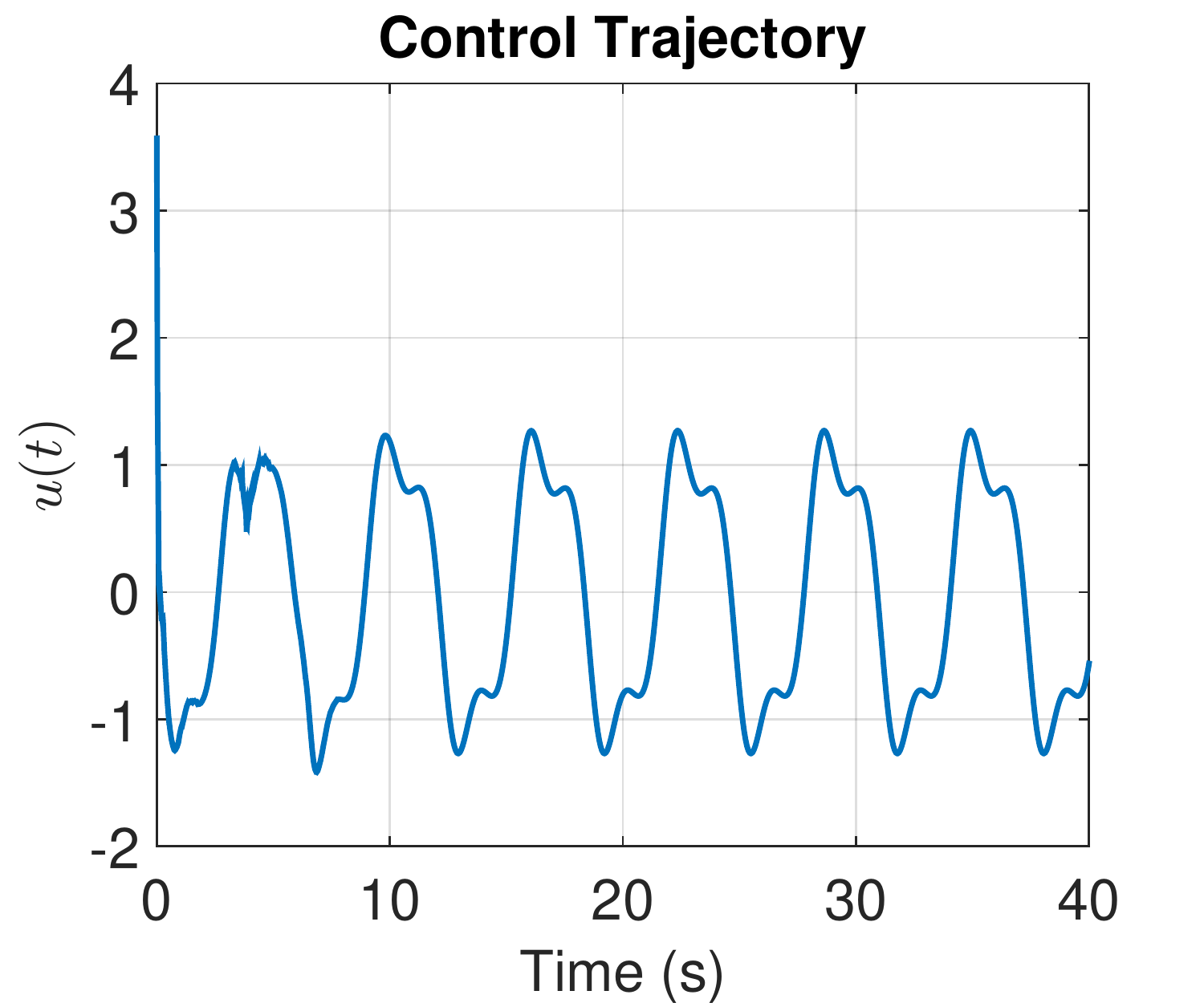}
\par\end{centering}

\noindent \centering{}\protect\caption{\label{fig:StaFSystTrajectories-1}Control signal generated using
the proposed method for the nonlinear system.}
\end{figure}

\begin{figure}[h]
\noindent \begin{centering}
\includegraphics[width=1\columnwidth]{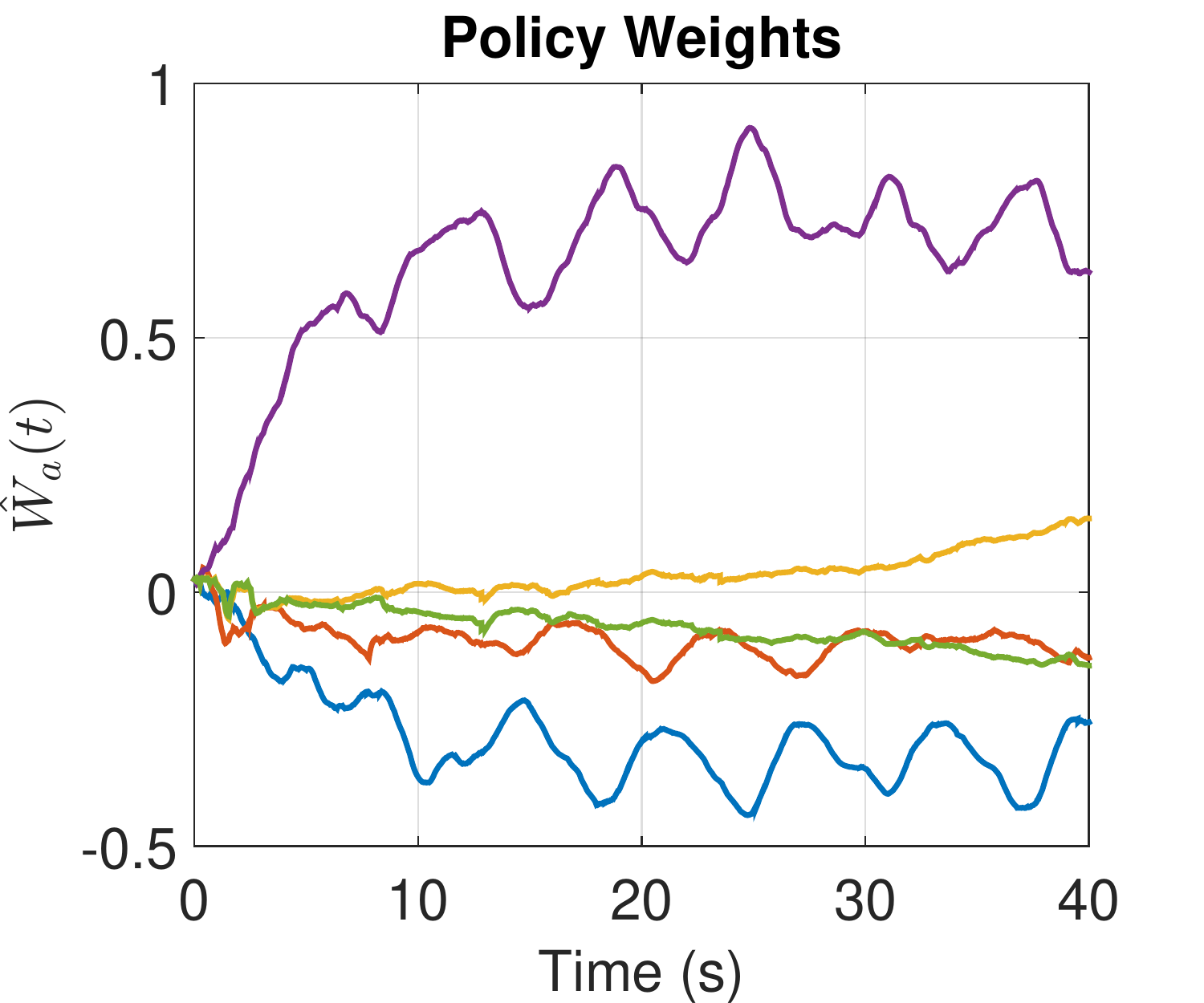}
\par\end{centering}

\noindent \centering{}\protect\caption{\label{fig:StaFWcWa}Policy weight trajectories generated using the
proposed method for the nonlinear system. The weights do not converge
to a steady-state value because the ideal weights are functions of
the time-varying system state. Since an analytical solution of the
optimal tracking problem is not available, weights cannot be compared
against their ideal values}
\end{figure}

\begin{figure}[h]
\noindent \begin{centering}
\includegraphics[width=1\columnwidth]{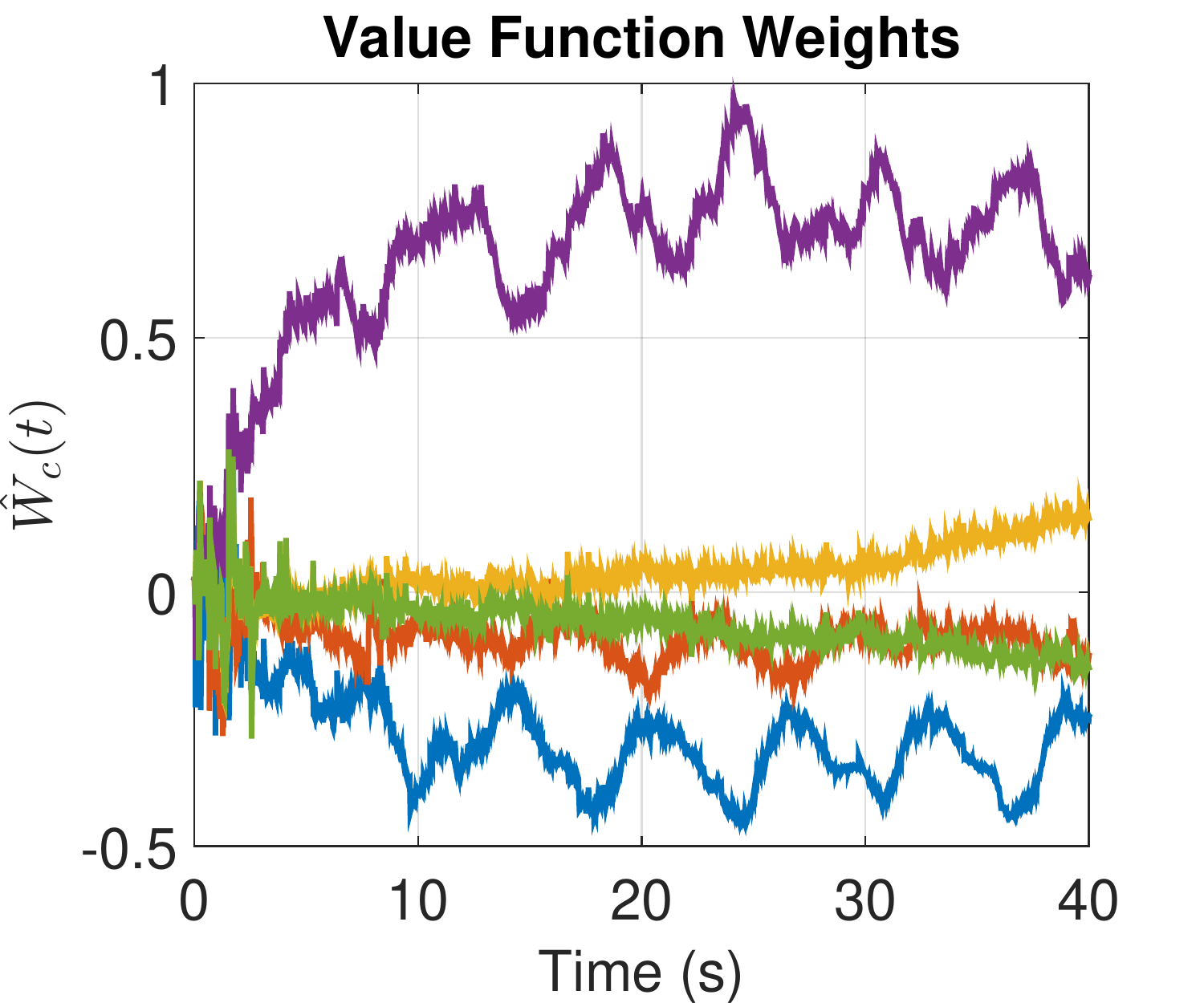}
\par\end{centering}

\noindent \centering{}\protect\caption{\label{fig:StaFWcWa-1}Value function weight trajectories generated
using the proposed method for the nonlinear system. The weights do
not converge to a steady-state value because the ideal weights are
functions of the time-varying system state. Since an analytical solution
of the optimal tracking problem is not available, weights cannot be
compared against their ideal values}
\end{figure}

\begin{figure}[h]
\noindent \begin{centering}
\includegraphics[width=1\columnwidth]{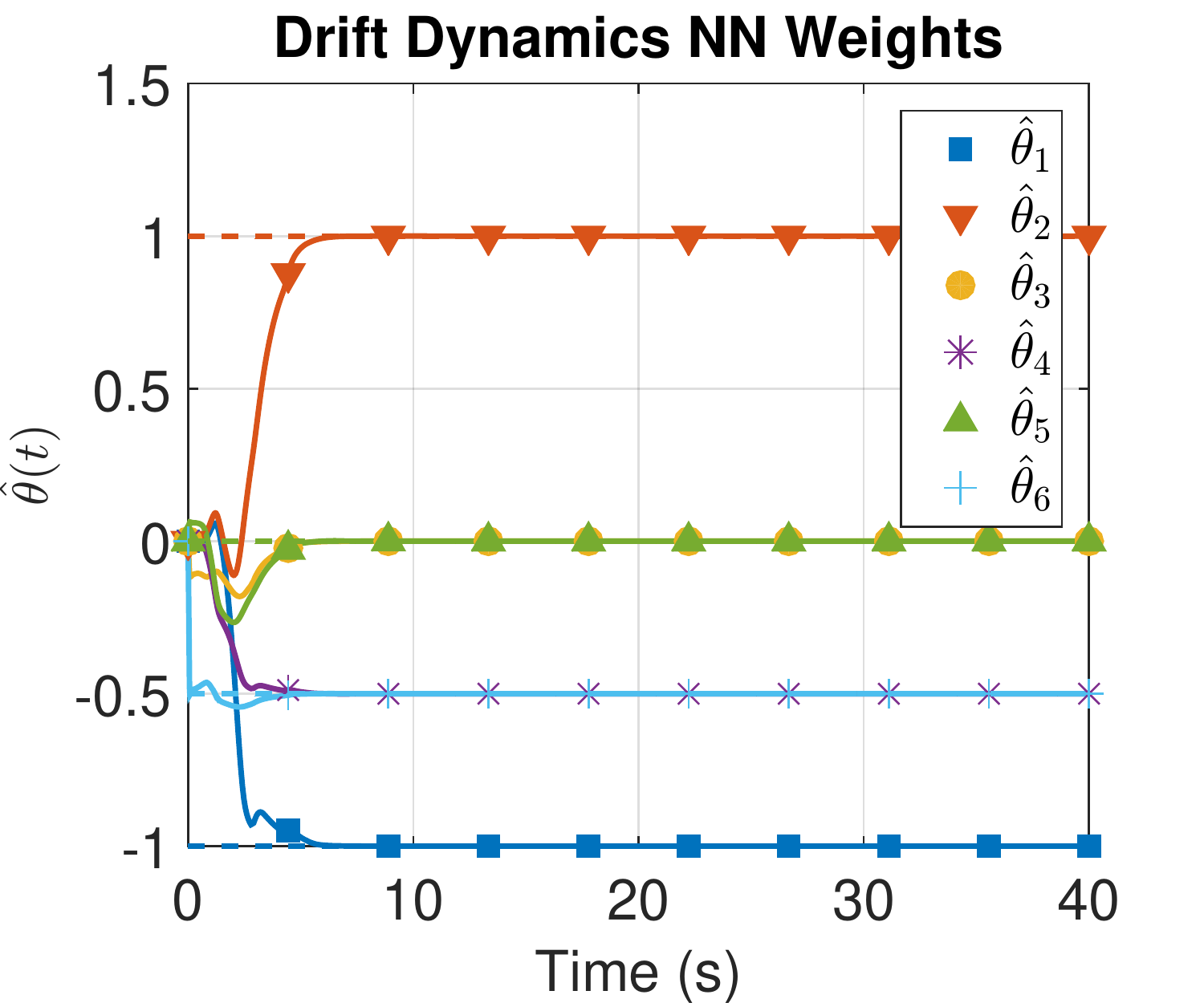}
\par\end{centering}

\noindent \centering{}\protect\caption{\label{fig:StaFTheta}Trajectories of the unknown parameters in the
system drift dynamics for the nonlinear system. The dotted lines represent
the true values of the parameters.}
\end{figure}

Figures \ref{fig:StaFTracking-error-trajectories} and \ref{fig:StaFSystTrajectories-1}
demonstrate that the controller remains bounded and the tracking error
is regulated to the origin. The NN weights are functions of the system
state $\zeta$. Since $\zeta$ converges to a periodic orbit, the
NN weights also converge to a periodic orbit (within the bounds of
the excitation introduced by the BE extrapolation signal), as demonstrated
in Figures \ref{fig:StaFWcWa} and \ref{fig:StaFWcWa-1}. Figure \ref{fig:StaFTheta}
demonstrates that the unknown parameters in the drift dynamics, represented
by solid lines, converge to their ideal values, represented by dashed
lines.

\subsection{Comparison}

\begin{table}
\begin{tabular}{>{\centering}p{0.2\columnwidth}>{\centering}p{0.2\columnwidth}>{\centering}p{0.2\columnwidth}>{\centering}p{0.2\columnwidth}}
\toprule 
Method & Running time (s) & Total cost & Steady-state RMS error\tabularnewline
\midrule
\midrule 
StaF kernels with single moving extrapolation points & 0.95 & 2.82 & $2.5\times10^{-3}$\tabularnewline
\midrule
\midrule 
Technique developed in \cite{Kamalapurkar.Walters.ea2013} & 2 & 1.83 & $6.15\times10^{-6}$\tabularnewline
\bottomrule
\end{tabular}

\protect\caption{\label{tab:ADPStaFSimCompare}Regulation simulation running times
for the developed technique and the technique in \cite{Kamalapurkar.Walters.ea2013}}
\end{table}
\begin{table}
\begin{tabular}{>{\centering}p{0.2\columnwidth}>{\centering}p{0.2\columnwidth}>{\centering}p{0.2\columnwidth}>{\centering}p{0.2\columnwidth}}
\toprule 
Method & Running time (s) & Total cost & Steady-state RMS error\tabularnewline
\midrule
\midrule 
StaF kernels with single moving extrapolation points & 15 & 6.38 & $2.13\times10^{-4}$\tabularnewline
\midrule
\midrule 
Technique developed in \cite{Kamalapurkar.Andrews.ea2014} & 103 & 3.1 & $2.7\times10^{-4}$\tabularnewline
\bottomrule
\end{tabular}

\protect\caption{\label{tab:ADPStaFSimCompare-1}Tracking simulation running times
for the developed technique and the technique in \cite{Kamalapurkar.Andrews.ea2014}}
\end{table}
The developed technique is compared with the model-based RL method
developed in \cite{Kamalapurkar.Walters.ea2013} for regulation and
\cite{Kamalapurkar.Andrews.ea2014} for tracking, respectively. Both
the simulations are performed in MATLAB$^{\circledR}$ SIMULINK$^{\circledR}$
at 1000 Hz on the same machine. The regulation simulations run for
10 seconds of simulated time, and the tracking simulations run for
40 seconds of simulated time. Tables \ref{tab:ADPStaFSimCompare}
and \ref{tab:ADPStaFSimCompare-1} show that the developed controller
requires significantly fewer computational resources than the controllers
from \cite{Kamalapurkar.Walters.ea2013} and \cite{Kamalapurkar.Andrews.ea2014}. 

Since the optimal solution for the regulation problem is known to
be quadratic, the model-based RL method from \cite{Kamalapurkar.Walters.ea2013}
is implemented using three quadratic basis functions. Since the basis
used is exact, the method from \cite{Kamalapurkar.Walters.ea2013}
yields a smaller steady-state error than the developed method, which
uses three inexact, but generic StaF kernels. For the tracking problem,
the method from \cite{Kamalapurkar.Andrews.ea2014} is implemented
using ten polynomial basis functions selected based on a trial-and-error
approach. The developed technique is implemented using five generic
StaF kernels. In this case, since the optimal solution is unknown,
both the methods use inexact basis functions, resulting in similar
steady-state errors. 

The two main advantages of StaF kernels are that they are universal,
in the sense that they can be used to approximate a large class of
value functions, and that they target local approximation, resulting
in a smaller number of required basis functions. However, the StaF
kernels trade optimality for universality and computational efficiency.
The kernels are inexact, and the weight estimates need to be continually
adjusted based on the system trajectory. Hence, as shown in Tables
\ref{tab:ADPStaFSimCompare} and \ref{tab:ADPStaFSimCompare-1}, the
developed technique results in a higher total cost than state-of-the-art
model-based RL techniques.

\section{Conclusion}

In this paper an infinite horizon optimal control problem is solved
using a new approximation methodology called the StaF kernel method.
Motivated by the fact that a smaller number of basis functions is
required to approximate functions on smaller domains, the StaF kernel
method aims to maintain good approximation of the value function over
a small neighborhood of the current state. Computational efficiency
of model-based RL is improved by allowing selection of fewer time-varying
extrapolation trajectories instead of a large number of autonomous
extrapolation functions. Simulation results are presented that solve
the infinite horizon optimal regulation and tracking problems online
for a two state system using only three and five basis functions,
respectively, via the StaF kernel method.

State-of-the-art solutions to solve infinite horizon optimal control
problems online aim to approximate the value function over the entire
operating domain. Since the approximate optimal policy is completely
determined by the value function estimate, state-of-the-art solutions
generate policies that are valid over the entire state space. Since
the StaF kernel method aims at maintaining local approximation of
the value function around the current system state, the StaF kernel
method lacks memory, in the sense that the information about the ideal
weights over a region of interest is lost when the state leaves the
region of interest. Thus, unlike existing techniques, the StaF method
generates a policy that is near-optimal only over a small neighborhood
of the origin. A memory-based modification to the StaF technique that
retains and reuses past information is a subject for future research.

\bibliographystyle{IEEEtran}
\bibliography{master,ncr,encr}

\end{document}